\documentclass[12pt,preprint]{aastex}

\slugcomment{accepted for publication in ApJ}

\shorttitle{SiO Maser Intensity Ratio}
\shortauthors{Nakashima, J. \& Deguchi, S.}

\begin{document}


\title{Correlation between Infrared Colors and Intensity Ratios\\ of SiO Maser Lines}


\author{Jun-ichi Nakashima\altaffilmark{1,2} and Shuji Deguchi\altaffilmark{3}}

\altaffiltext{1}{Academia Sinica Institute of Astronomy and Astrophysics, P.O. Box 23-141, Taipei 10617, Taiwan; email(JN): junichi@asiaa.sinica.edu.tw}

\altaffiltext{2}{Department of Astronomy, University of Illinois at Urbana-Champaign,
1002 W. Green St., Urbana, IL 61801}

\altaffiltext{3}{Nobeyama Radio Observatory, National Astronomical Observatory, Minamimaki, Minamisaku, Nagano 384-1305, JAPAN; deguchi@nro.nao.ac.jp}


\begin{abstract}
We present the results of SiO millimeter-line observations of a sample of known SiO maser sources covering a wide dust-temperature range. A cold part of the sample was selected from the SiO maser sources found in our recent SiO maser survey of cold dusty objects. The aim of the present research is to investigate the causes of the correlation between infrared colors and SiO maser intensity ratios among different transition lines. In particular, the correlation between infrared colors and SiO maser intensity ratio among the $J=1$$-$0 $v=$1, 2, and 3 lines are mainly concerned in this paper. We observed in total 75 SiO maser sources with the Nobeyama 45m telescope quasi-simultaneously in the SiO $J=1$$-$0 $v=0$, 1, 2, 3, 4 and $J=2$$-$1 $v=1$, 2 lines. We also observed the sample in the $^{29}$SiO $J=1$$-$0~$v=0$ and $J=2$$-$1 $v=0$, and $^{30}$SiO $J=1$$-$0 $v=0$ lines, and the H$_2$O 6$_{1,6}$$-$5$_{2,3}$ line. As reported in previous papers, we confirmed that the intensity ratios of the SiO $J=1$$-$0 $v=2$ to $v=1$ lines clearly correlate with infrared colors. In addition, we found possible correlation between infrared colors and the intensity ratios of the SiO $J=1$$-$0 $v=3$ to $v=1\&2$ lines. Two overlap lines of H$_2$O (i.e., $11_{6,6}$~$\nu_{2}=1$$\rightarrow$$12_{7,5}$~$\nu_{2}=0$ and $5_{0,5}$~$\nu_{2}=2$$\rightarrow$$6_{3,4}$~$\nu_{2}=1$) might explain these correlation if these overlap lines become stronger with increase of infrared colors, although the phenomena also might be explained by more fundamental ways if we take into account the variation of opacity from object to object.
\end{abstract}


\keywords{masers ---
stars: AGB and post-AGB ---
stars: late-type ---
stars: mass loss ---
stars: statistics}


\section{Introduction}
Since the discovery of the SiO maser emission toward Orion IRc2 \citep{sny74}, roughly 2300 SiO maser sources have been so far found in the sky \citep{deg07}. Although the SiO maser emission has been first detected toward a star-forming region, ironically most pursuant SiO maser sources have been identified as evolved stars with a thick dust envelope \citep{kai75,rei81}. The SiO maser emission is nowadays applied to study a wide variety of astrophysical problems from circumstellar kinematics of evolved stars \citep[e.g.,][]{ima99,dia03,cot04,yi05} to the galactic dynamics \citep[e.g.,][]{jia96,izu99,ita01,miy01,mes02,nak03a,deg04,fuj06}. However, we still do not fully understand the fundamental pumping mechanism of the SiO masers \citep[e.g.,][]{buj94,doe95}.

An important problem in the studies on the SiO masers was that SiO maser sources ever known were considerably biased. Specifically, the dust (effective) temperature of known SiO maser sources, which was calculated from mid-infrared flux densities (such as the IRAS and MSX flux densities), was limited roughly in a range of 250 K $\lesssim T_{\rm dust} \lesssim 2000$ K. This is because the previous SiO maser surveys have mainly aimed to study the galactic dynamics through the motion of SiO maser sources. To do such investigation, high detection rates of the SiO maser search and homogeneity of the sample are essential. Therefore, the majority of the previous SiO maser surveys were made in a specific dust-temperature range, in which the detection rate of the SiO maser search maximizes. Consequently a non-negligible number of potential SiO maser sources (especially with a low dust-temperature) have been slipped from the previous SiO maser surveys. 

\citet{nym93} first realized the importance of SiO maser sources exhibiting a low dust-temperature. They investigated how SiO maser emission behaves in a low dust-temperature range by observing OH/IR stars in the SiO $J=1$$-$0 $v=1\&2$ and $J=$2--1 $v=1$ lines. The OH/IR stars often exhibit a low dust-temperature less than $T_{\rm dust} = 250$ K. In their observation cold objects clearly show a larger intensity ratio of the SiO $J=1$$-$0 $v=2$ to $v=1$ lines. Both collisional and radiative schemes cannot fully explain this observational properties of the SiO masers \citep{buj94,doe95}. \citet{nym93} suggested that an infrared H$_2$O line ($11_{6,6}$~$\nu_{2}=1$$\rightarrow$$12_{7,5}$~$\nu_{2}=0$) overlapping with the SiO $J=0$ $v=1$$\rightarrow$$J=1$ $v=2$ transition might play an important role. This overlap line of H$_2$O was first introduced by \citet{olo81,olo85} to explain the anomalous intensity of the SiO $J=2$$-$1 $v=2$ line. However, in early 1990s the number of cold SiO maser sources (like OH/IR stars) was quite limited, and it was difficult to statistically futher investigate the relation between infrared colors and intensity ratios of SiO maser lines.

\citet{nak03b} recently extended the Nyman's study by surveying the SiO maser emission in cold, dusty IRAS sources exhibiting low dust temperature less than 250 K. They found roughly 40 new SiO maser sources in the cold dusty objects, and in conjunction with the results of another SiO maser survey of relatively warmer IRAS objects \citep{nak03a}, they clearly demonstrated that the intensity ratio of the SiO $J=1$$-$0 $v=2$ to $v=1$ lines increases in inversely proportional to the dust temperature. \citet{nak03b} again suggested that the overlap line of H$_2$O might explain this correlation if the overlap line becomes stronger with decrease of the dust temperature. To consider further this problems, we need to confirm whether properties of the SiO lines other than $J=1$$-$0 $v=1$ and 2 lines are consistent with the existence of the H$_2$O overlap line.

In this paper we present the result of quasi-simultaneous observations in the multiple different SiO rotational lines with the Nobeyama 45m telescope. The aim of this observational research is to investigate the correlation between dust-temperature (infrared colors) and SiO maser intensity ratios among different transition lines. The outline of the paper is as follows. In Sect 2 details of sample selection, observation and data reduction are presented. In Sect 3, the observational properties of the SiO $J=1$$-$0 $v=1$, 2 and 3 lines are discussed, and those of other observed lines are also briefly mentioned. In Sect 4, we discuss the possible explanations of the correlations between infrared colors and the SiO maser intensity ratios. Finally, the results of the present research are summarized in Sect 5.

\section{Observation}
\subsection{Sample}
The observing targets were selected from \citet{nak03a,nak03b} and the Nobeyama SiO maser source catalog (Gorny et al. in preparation) in terms of the IRAS colors and flux densities. The targets are distributed roughly in the right ascension range between $18^{\rm h}$ and $22^{\rm h}$, because the cold SiO maser sources found by \citet{nak03b} are distributed roughly in this range. We selected the observing targets basically in order of the brightness at $\lambda =12$~$\mu$m, but we also paid attention to the source distribution in the IRAS two-color diagram (see, Figure 1) so that the observing targets continuously cover the entire color range. In Figure 1 the observing targets mainly occupy the regions I, II, IIIab, VII and IV, in which oxygen-rich asymptotic giant branch (AGB) stars are dominant \citep{van88}, but some SiO maser sources still exist in the region VIII and V, in which presumably post-AGB stars and/or AGB stars with a large mass-loss rate are lying. For example, OH 231.8+4.2, W43A and IRAS 19312+1950 are located in the region VIII or V \citep{mor87,nak00a,ima02,nak04a,nak04b,nak05}. Finally we selected 75 observing targets. The selected targets are listed in Table 1 along with infrared fluxes taken from the 2MASS, IRAS and MSX point source catalogs \citep{joi94,ega99,cut03}. For the sake of later discussions, luminosity distances to the targets are also listed in Table 1. This luminosity distances were calculated from bolometric fluxes obtained by integrating infrared fluxes given in Table 1 on the assumption of 8000~$L_{\odot}$, which is a typical absolute luminosity of stellar SiO maser sources \citep[e.g.,][]{nak00}. Since details about the calculation of the luminosity distance to SiO maser sources are found elsewhere \citep[e.g.,][]{nak00,deg02}, we do not repeat it here. We discuss the uncertainty of the luminosity distance later in Sect 3.2. For known red supergiants, the distances are not given in Table 1, because the absolute luminosity of these objects are significantly different from AGB stars.

\subsection{Details of observations and data reductions}

SiO line observations with the Nobeyama 45m telescope \citep{kai85} were made in two separated periods: May 11--19, 2004 and February 15--19, 2006. In the first period we observed, in total, 38 objects with the cooled SIS-mixer receivers, S40 and S100. The frequency coverage of S40 and S100 was roughly 500 MHz. The observed SiO transitions in the first period were $J=1$$-$0 $v=$1, 2, 3 and $J=2$$-$1 $v=$1, 2. We also observed in the $^{29}$SiO $J=1$$-$0 $v=$0 and $J=2$$-$1 $v=$0 lines. These lines were observed with different frequency settings in different days, because the rest frequencies of the lines could not be simultaneously covered by the frequency coverage of S40 and S100. Maser lines are known to exhibit weak polarization \citep[typical fractional polarizations are 5\%--30\%: see, e.g.,][]{bar85}. Therefore, in the first period we observed each object at an almost same local sidereal time on every different day to keep a rotation angle of the receiver within a certain small range (deviation of the observing time from day to day was less than $\pm$1.5 hours). In addition, in the first period we observed 27 objects in the H$_2$O maser line at 22 GHz ($6_{1,6}$$-$5$_{2,3}$) as a backup observation under rainy/heavy cloudy condition. In this H$_2$O maser observations, we used the HEMT receiver, H22. 

In the second period we observed, in total, 53 objects with the HEMT receiver H40 with a frequency coverage of roughly 2.0 GHz. The observed transitions in the second period were SiO $J=1$$-$0 $v=$0, 1, 2, 3, 4, $^{29}$SiO $J=1$$-$0 $v=$0 and $^{30}$SiO $J=1$$-$0 $v=$0. The rest frequencies of these lines were simultaneously covered by the wide frequency coverage of the H40 receiver. 

Throughout the first and second periods we used two different acousto-optic spectrometers, AOS-H and AOS-W. Both AOS-H and AOS-W have 2048 frequency channels. The frequency coverage of AOS-H and AOS-W was 40 and 250 MHz, respectively. We used 16 spectrometers at a time (AOS-H$\times8$ and AOH-W$\times8$) to simultaneously cover multiple different lines. We used AOS-H for the lines in the 43 and 22 GHz bands (i.e., SiO $J=1$$-$0 $v=$0, 1, 2, 3, 4, $^{29}$SiO $J=1$$-$0 $v=$0, $^{30}$SiO $J=1$$-$0 $v=$0 and H$_2$O $6_{1,6}$$-$5$_{2,3}$), and used AOS-W for the lines in the 86 GHz band (i.e., SiO $J=2$$-$1 $v=$1, 2 and $^{29}$SiO $J=2$$-$1 $v=$0). This was because the noise level at 86 GHz was higher than that at 43 and 22 GHz. Rest frequencies of the observed lines were assigned to the center of the frequency coverage of spectrometers. Velocity coverage of AOS-H at 43 and 22 GHz was roughly 280 and 540 km s$^{-1}$, respectively, and that of AOS-W at 86 GHz was 870 km s$^{-1}$. System temperatures of S40 and H22 were roughly 180--220 K, and those of S100 and H40 were roughly 250--350 K (depending on the telescope position, observing frequency and weather condition). The rest-frequencies of the observed lines were basically taken from \citet{lov92}, but that only of the SiO $J=1$$-$0 $v=4$ line was taken from \citet{sny86}. The observations were made with a position switching mode, and the off-position was taken at 5$'$ east of the target positions in the azimuth direction. The telescope pointing was checked every 1--2 hours with a 5-point cross mapping of strong SiO maser sources. Several targets in the present sample were used for the telescope pointing (for example, V1111 Oph [=IRAS 18349+1023], $\chi$ Cyg [=IRAS 19486+3247]). The pointing accuracy of the telescope was, in most cases, better than 10$''$, but depending on wind speed. The beam sizes at 22, 43 and 86 GHz were 73$''$, 38$''$ and 18$''$, respectively. The aperture efficiency at 22, 43, and 86 GHz were 63\%, 60\% and 44\%, respectively. The conversion factors from K to Jy at 22, 43 and 86 GHz were 2.98, 2.89 and 4.00 Jy/K, respectively. 

We observed at the MSX positions of the targets whenever those are available, because the angular resolution of the MSX survey is roughly 30 times better than that of the IRAS survey \citep{pri01}, and also because the accuracy of the IRAS positions are occasionally insufficient even for the single dish observations especially in the case of infrared faint objects \citep{deg01,nak03b}. The present sample includes some objects lying out of the MSX survey region, and the MSX positions are not available there. However, the accuracy of the IRAS positions of these objects is considered to be sufficient enough for the present observations, because all the sources out of the MSX region are quite bright at mid-infrared wavelengths. We reduced raw data using the reduction software, NEWSTAR, developed by the Nobeyama Radio Observatory \citep{ike01}. The reduction procedure included flagging out bad data, integrating the data in time, and removing a slope in the base line by least-square fitting of the first order polynomial.

\section{Results}

A statistical summary of the observations is given in Table 2, and spectra of all detected lines are presented in Figure 2. Results of the observations are presented in Table 3, including IRAS names of observing targets, observed transitions, radial velocities, peak intensities, velocity-integrated intensities, upper limits of the velocity-integrated intensities, rms noise levels and observing dates. (Figure 2 and Table 3 are given only in electric form.) The radial velocities are the averages of radial velocities at the first and last 3$\sigma$ channels. Although the velocity at the intensity peak is often used as a velocity of SiO maser sources, this is not appropriate for some objects in the present sample, for example, IRAS 19254+1631 and IRAS 18349+1023, IRAS 23041+1016 (see, Figure 2). The upper limits of the integrated intensities are defined by the following formula: $S_{\rm upper}=3 \sigma\ n^{1/2}\ \Delta v$, where $\sigma$ is the rms of flux density, $\Delta v$ is the effective velocity resolution, and $n$ is the number of channels. To calculate the upper limits we assumed that the line width is 3 km s$^{-1}$. The number of observations given in Table 2 is larger than the total number of the targets for some transitions, because we have repeated the observation of some targets so that we can check the time variation of the maser intensity ratios. In following subsections, we first discuss the properties of the SiO $J=1$$-$0 $v=1$, 2 and 3 lines, in which we detected an enough number of objects for statistical analysis. Then we briefly summarize the properties of the other lines.

\subsection{Properties of the SiO $J=1$$-$$0$ $v=1$, $2$ and $3$ lines}
\subsubsection{Infrared colors versus intensity ratios of SiO maser lines}
Figure 3 shows the relations between infrared colors and intensity ratios among the SiO maser lines. The line intensities used to calculate the intensity ratios are velocity-integrated intensities (presented in the 5th column in Table 3). The intensity ratios were calculated with the data obtained within a single observing period, because the intensity of SiO maser lines exhibits large time-variation with a time-scale of a few hundred days \citep[see, e.g.,][]{gom86,lee94,kam05}. Here we omit to present similar diagrams with peak intensities, because the peak intensities give quite similar properties with the velocity-integrated intensities. The values of the peak intensity itself are given in the 4th column in Table 3. W51 has been excluded from Figure 3, because no reliable infrared fluxes of the core emitting the SiO maser line are available, and also because \citet{nak03b} have already comparatively discussed the properties of SiO maser emission between young stellar objects (including W51) and evolved stars.

In Figure 3 we applied three different infrared colors: $\log(F_{25}/F_{12})$, $\log(F_{\rm e}/F_{\rm c})$ and $H$$-$$K$, where $F_{25}$ and $F_{12}$ are the IRAS flux densities at $\lambda=25$ and 12~$\mu$m, $F_{\rm e}$ and $F_{\rm c}$ are the MSX flux densities at $\lambda=21.3$ and 12.13~$\mu$m, and the $H$$-$$K$ color was calculated with the 2MASS $H$- and $K$-bands magnitudes. Correlation coefficients were calculated for each pair of the intensity ratios and infrared colors, and the obtained values of the coefficients are given in the lower-right corners of each panel in Figure 3. 

As stated above, we observed 9 objects both in the first and second periods to check the time-variation of the line intensity ratios. The difference of the intensity ratios of the SiO $J=1$$-$0 $v=2$ to $v=1$ lines between two observing periods reaches up to 0.44 in a logarithmic scale (corresponding to factor 2.8). The intensity ratios of the SiO $J=1$$-$0 $v=3$ to $v=2$ lines and of the SiO $J=1$$-$0 $v=3$ to $v=1$ lines exhibit, more or less, similar differences with that of the ratio of the SiO $J=1$$-$0 $v=2$ to $v=1$ lines. The maximum differences of factor 2.8 are consistent with previous monitoring observations of M-type miras. For example, \citet{mci06a,mci06b} monitored Mira ($o$ Cet) in the SiO $J=1$$-$0 $v=1$ and 2 lines, and showed that that the maximum difference of the intensity ratio of the SiO $J=1$$-$0 $v=2$ to $v=1$ lines reaches up to roughly factor 2.5 in a single pulsation period. In Figure 3 we plotted both two data points obtained in the different observing periods for the 9 repeated objects.

In the panel A in Figure 3 we can clearly confirm the positive correlation between the intensity ratio of the SiO $J=1$$-$0 $v=2$ to $v=1$ lines and the $\log(F_{25}/F_{12})$ color as first reported by \citet{nak03b}. The intensity ratio of the SiO $J=1$$-$0 $v=2$ to $v=1$ lines also clearly exhibits correlation with the $\log(F_{\rm e}/F_{\rm c})$ color (see, panel D). In the panels A and D, the two downward triangles (representing upper limits of the intensity ratio) lying at $\log(F_{25}/F_{12}) \sim 0.6$ and $\log(F_{\rm e}/F_{\rm c}) \sim 0.5$ are placed clearly below the dashed line. This feature is consistent with the breaking-down of the correlation at $\log(F_{25}/F_{12}) \sim 0.5$ suggested by \citet{nak03b}. Incidentally, the color of  $\log(F_{25}/F_{12})=0.5$ corresponds to the boundary between the distributions of AGB and post-AGB stars in the $\log(F_{25}/F_{12})$ color \citep{van88,nak03b}. Similar correlation is also seen in panel G showing the relation between the intensity ratio of the SiO $J=1$$-$0 $v=2$ to $v=1$ lines and the $\log(F_{25}/F_{12})$ color.

Interestingly, the intensity ratios of the $J=1$$-$0 $v=3$ to $v=2$ lines and of the $J=1$$-$0 $v=3$ to $v=1$ lines seem to also correlate with the $\log(F_{25}/F_{12})$ color (see, panels B and C) even though the correlation coefficients are slightly smaller than that of panel A. On the other hand, the correlation of these ratios with the $\log(F_{\rm e}/F_{\rm c})$ color is unclear (see, panels E and F), and the correlation coefficients are in fact close to 0. The intensity ratios of the $J=1$$-$0 $v=3$ to $v=2$ lines and of the $J=1$$-$0 $v=3$ to $v=1$ lines seem to weakly correlate with the $H$$-$$K$ color (see, panels H and I) even though the correlation coefficients are small ($-$0.03 and 0.23). Some exceptional data points presumably cause this small correlation coefficients: in panel H, ($H$$-$$K$, $\log (I_{2}/I_{1})$)$=(0.2, 0.5), (6.5, -1.35), (6.5, -1.7)$, and in panel I, ($H$$-$$K$, $\log (I_{2}/I_{1})$)$=(0.2, 0.5), (6.5, -1.0), (6.5, -1.5)$.

\subsubsection{Absolute intensity versus infrared colors}

Figure 4 shows the relations between infrared colors and absolute intensities of the SiO maser lines. The intensity of the SiO maser lines is standardized at the distance of 1~kpc using the luminosity distances given in Table 1. The left column of Figure 4 show the relations between the $\log(F_{25}/ F_{12})$ color and the absolute intensity of the SiO $J=1$$-$0 $v=1$, 2 and 3 lines. A notable feature seen in these panels is that the SiO maser absolute intensities undoubtedly correlate with the $\log(F_{25}/ F_{12})$ color. Another clear feature is that the higher the vibrational transitions, the steeper the inclination of the dashed lines, which represent the results of least-square-fitting of a first order polynomial (the inclinations of the dashed lines are given in the lower-right corners of each panel along with statistical uncertainty). This tendency is consistent with the correlation seen in Figure 3.

In the panel A in Figure 4, the values of the absolute intensity of SiO maser emission seem to maximize at $\log(F_{25}/ F_{12}) \sim 0.5$, and the values tend to decrease with increase of the color in the red region above $\log(F_{25}/ F_{12}) = 0.5$. The $\log(F_{25}/ F_{12})$ color of 0.5 corresponds the boundary between distributions of AGB and post-AGB stars in the $\log(F_{25}/ F_{12})$ color as stated in Sect 3.1. In fact, the panel A in Figure 3 \citep[and Figure 8 in][]{nak03b} shows a sudden change of the feature at $\log(F_{25}/ F_{12}) \sim 0.5$. No such change is seen in the panels D and G in Figure 4, simply because the SiO $J=1$$-$0 $v=2$ and 3 lines have not been detected above $\log(F_{25}/ F_{12}) = 0.5$ in the present observation. The middle column of Figure 4 similarly show the relations between the $\log(F_{\rm e}/ F_{\rm c})$ color and the absolute intensity of the SiO $J=1$$-$0 $v=1$, 2 and 3 lines. Features seen in these panels are basically same with those seen in the left column except that the number of detections in the blue region (i.e., $\log(F_{\rm e}/ F_{\rm c}) \lesssim 0$) is small. 

The right column of Figure 4 shows the relation between the $H$$-$$K$ color and the absolute intensity of the SiO $J=1$$-$0 $v=1$, 2 and 3 lines. The feature seen in the right column is somewhat different from those seen in the left and middle columns, exhibiting a distribution showing a triangle-like shape. This is presumably because a non-negligible number of SiO maser sources with a very red IRAS/MSX color somehow exhibit a relatively blue $H$$-$$K$ color.

Here, we ought to discuss the uncertainty of the ``absolute'' intensity of the SiO maser lines, because the data points in Figures 3 and 4 exhibits nonnegligible scatter, and also because the degree of the scatter is critical in further discussions and interpretations in later sections. As mentioned in Sect 2.1, the luminosity distances given in Table 1 were calculated on the assumption that the absolute luminosity of the targets is 8000~$L_{\odot}$, which represents a typical absolute luminosity of AGB stars \citep[e.g.,][]{vas93}. The present sample, in principle, can include some red supergiants in terms of infrared colors [for example, the $\log(F_{25}/F_{12})$ color of a typical supergiant, VY CMa, is $-0.17$]. However the majority of the sample must be AGB stars, because the SiO maser emission in the red supergiants is quite rare and weak \citep{alc90}. Besides, recent observations revealed that the red supergiants emitting SiO maser lines are preferably lying in a young star cluster and rarely exist in isolated environments \citep{nak06}. In fact, according to the MSX and 2MASS images, almost all of the objects in the present sample are clearly isolated, and do not show any evidences for lying in a star cluster. Furthermore, \citet{nak03a,nak03b} demonstrated, by making use of the luminosity distances of SiO maser sources obtained under the assumption of 8000~$L_{\odot}$, that kinematical properties of SiO maser sources are consistent with those of the gas component in the Galaxy. Thus, in our opinion, the assumption that the all objects in the sample are AGB stars would be appropriate. To make sure, the known red supergiant, IRAS 07209$-$2540 (VY CMa), was excluded from Figures 3 and 4. 

Under the assumption of the absolute luminosity of 8000~$L_{\odot}$, relative uncertainty of the luminosity distance is considered to be less than roughly 30\%. There are a couple of observational evidences for this relative uncertainty of the luminosity distance. For example, \citet{deg01} determined luminosity distances to the AGB stars in the Galactic center star cluster under the assumption of 8000~$L_{\odot}$, and they found that the standard deviation of the luminosity distances to the Galactic center AGB stars are less than 30\%. For another evidence, distances obtained with the period-luminosity relation of miras \citep{nak00} correspond with the luminosity distances obtained under the assumption of 8000~$L_{\odot}$ within a relative uncertainty of roughly 30\%.

Another important factor affecting on the SiO maser absolute intensity is the time variation of the SiO maser intensity caused by the pulsation of AGB stars. Although there is no successful models fully explaining the physical relation between the pulsation and the time variation of the SiO maser intensity, a tight correlation between the optical/infrared and SiO maser intensities \citep[see, e.g.,][]{alc99,par04,kan06} strongly suggests that the time variation of the SiO masers somehow connects to the pulsation of stars. The intensity of the SiO maser lines has been monitored over multiple pulsation periods by several authors in the $J=1$$-$0 $v=1$ and 2 transitions \citep[see, e.g.,][]{gom86,kam05,mci06a,mci06b}. According to these monitoring observations, the intensity of the SiO $J=1$$-$0 $v=1$ and 2 lines changed in factor of 20--50 over one pulsation period. [\citet{mci06a} monitored Mira ($o$ Cet) in the SiO $J=1$$-$0 $v=3$ line, but unfortunately the number of observations was insufficient to discuss the amplitude of the maser line intensity.] Thus, the scatter of data points seen in Figure 4 should be dominantly caused by the time variation of the SiO maser emission due to stellar pulsation rather than the uncertainty of the distance determination. In fact, the 30 \% uncertainty of the luminosity distance changes the absolute intensity of the SiO maser emission only between in factor of 0.5 and 2.0.

\subsubsection{SiO maser intensity versus 8 $\mu$m absolute flux}

A possible reason for the correlation seen in Figure 4 (especially in the left column) is that the energy input to the SiO maser region increases with the infrared colors. To confirm this possibility, in Figure 5 we plotted the 8~$\mu$m absolute flux densities as a function of the infrared colors. The values of the 8~$\mu$m flux densities were taken from the MSX point source catalog [the MSX $a$$-$band flux (center wavelength is 8.28~$\mu$m) was used]. If we rely on the radiative scheme the 8~$\mu$m flux should well represent the energy input to the SiO maser region, because the $\lambda=8$~$\mu$m corresponds to the $\Delta v=1$ SiO transition \citep{deg76,buj81,lan84}. In Figure 5 the 8~$\mu$m flux densities are standardized at the distance of 1~kpc using the luminosity distances given in Table 1. The distribution of the data points seen in Figure 5 is, in fact, strikingly similar with those seen in Figure 4, supporting that the 8~$\mu$m absolute flux tightly correlates with the SiO maser intensity. 

For the better understanding of the correlation between the 8~$\mu$m absolute flux and the SiO maser intensity, in Figure 6 we plotted the absolute SiO maser intensity as a function of the 8~$\mu$m absolute flux density. In Figure 6, all the three SiO maser lines (i.e., $J=1$$-$0 $v=1$, 2 and 3) clearly show a linear correlation with the 8~$\mu$m absolute flux density, and this result is consistent with a radiatively pumped saturated maser. This correlation between the SiO maser absolute intensity and 8~$\mu$m absolute flux density has been first reported by \citet{buj87} in the SiO $J=1$$-$0 $v=1$ and 2 lines; \citet{jia02} also confirmed the correlation by making use of the MSX data archive. The present result shows somewhat larger scatter comparing with the Bujarrabal's result. This is due mainly to the time variation of the SiO maser intensity caused by the stellar pulsation [Incidentally, \citet{buj87} observed their targets at the almost same pulsation phase by making use of their mid-infrared monitoring observations]. Nevertheless, the correlation between the 8~$\mu$m absolute flux and the SiO maser absolute intensity still leaves no doubt. The SiO maser absolute intensities also correlate with infrared continuum at other wavelengths, for example $\lambda=12$ and 25~$\mu$m, but the correlation coefficient maximizes at $\lambda=8$~$\mu$m; this result is also consistent with \citet{buj87}. 

Another notable feature in Figure 6 is that the distribution of the data points varies with the infrared colors. In Figure 6 the red filled circles and blue diamonds represent the objects with the $\log(F_{25}/ F_{12})$ color above and below 0.2, respectively. The cold objects represented by the red circles, in average, exhibit larger values both of the 8~$\mu$m absolute flux and SiO maser absolute intensity than the warm objects represented by the blue diamonds, but the both cold and warm objects (red circle and blue diamonds) are still lying in the same straight line. This result implies that the properties of the SiO maser emission of both cold and warm objects are consistent with the radiative scheme. In addition, the red circles in the upper panel of Figure 6 are distributed in a somewhat larger area comparing with those in the middle and lower panels. This is because some very red objects ($\log (F_{25}/F_{12}) \gtrsim 0.5$) exhibit weak SiO maser intensities as seen in the panel A in Figure 4.

\subsection{Properties of the other lines and line profiles}

In Figure 7 we show the relation between the $\log(F_{25}/F_{12})$ color and the SiO maser intensity ratio of the $J=2$$-$1 $v=1$ to $J=1$$-$0 $v=1$ lines. In Figure 7 the results of \citet{nym93} are also presented for comparison as the blue data points. The present observation, represented by the red data points, seems to be consistent with the Nyman's results. Mojority of the data points exhibit the intensity ratio around $-0.5$, but interestingly some data points exhibit small ratios in the red color larger than $\log(F_{25}/F_{12})=0.2$.

The absolute intensities of the SiO $J=1$$-$0 $v=0$ and $^{29}$SiO $J=1$$-$0 $v=0$ lines seem to correlate with infrared colors as well as the $J=1$$-$0 $v=1$, 2 and 3 lines. We also check the correlation between infrared color and the absolute intensity of the H$_2$O 6$_{1,6}$$-$5$_{2,3}$ line, but there is no correlation.

Some authors \citep[e.g., ][]{nym93} have pointed out that the profile of the $J=1$$-$0 $v=1$ line is very similar with that of the $J=1$$-$0 $v=2$ line, but is different from that of the $J=2$$-$1 $v=1$ line. However, a part of the present results contradict the previous observations. For example, IRAS 09448+1139, IRAS 18545+1040 and IRAS 18592+1455 clearly show different line profiles between the $J=1$$-$0 $v=1$ and 2 lines, and IRAS 19192+0922 and IRAS 20491+4236 show a very similar profile between the $J=1$$-$0 $v=1$ and $J=2$$-$1 $v=1$ lines (see, Figure 2). In addition we cannot find any relation between the infrared colors and similarity of the line profile, at least, by our eye inspection.


\section{Discussion}

In this section we go back to the main question in this paper: how can we explain the correlation between the infrared colors and the intensity ratios among the SiO maser lines (in particular among the $v=1$, 2 and 3 lines at 43 GHz)? One possible explanation is to introduce the overlap line of H$_2$O ($11_{6,6}$~$\nu_{2}=1$$\rightarrow$$12_{7,5}$~$\nu_{2}=0$), which has been first suggested by \citet{olo81,olo85} to explain the anomalous, weak intensity of the SiO $J=2$$-$1 $v=2$ line in oxygen-rich (O-rich) stars. A recent theoretical calculation also predicts that this overlap line of H$_2$O may explain the unexpected observational results found in VLBI maps \citep{sor04}. This H$_2$O line overlaps with the SiO $J=0$ $v=1$$\rightarrow$$J=1$ $v=2$ transition with a velocity difference of 1 km s$^{-1}$. With this line overlap, the $J=1$ $v=2$ level is overpopulated, and the weakness of the SiO $J=2$$-$1 $v=2$ line is explained by this overpopulation. The overpopulation at the $J=1$ $v=2$ level is also consistent with the strong intensity of the $J=1$$-$0 $v=2$ line. Thus, the correlation between the infrared colors and the intensity ratio of the SiO $J=1$$-$0 $v=2$ to $v=1$ lines may be explained if this overlap line of H$_2$O becomes stronger with increase of the infrared colors.

One problem in this interpretation is that the intensity ratios of the SiO $J=1$$-$0 $v=3$ to $v=1\&2$ lines cannot be explained only by the H$_2$O $11_{6,6}$~$\nu_{2}=1$$\rightarrow$$12_{7,5}$~$\nu_{2}=0$ line. However \citet{cho07} recently reported an interesting detection of the SiO $J=2$$-$1 $v=3$ line toward an S-type star, $\chi$ Cyg. They also confirmed that the SiO $J=2$$-$1 $v=3$ line is weak in O-rich stars. The S-type stars have almost same amount of oxygen and carbon atoms, and consequently they have few H$_2$O molecules in its envelope. These results potentially suggest that another overlap line of H$_2$O affects on the population distribution of SiO in O-rich stars, and \citet{cho07} have suggested that the H$_2$O $5_{0,5}$~$\nu_{2}=2$$\rightarrow$$6_{3,4}$~$\nu_{2}=1$ line overlapping with the SiO $J=0$ $v=2$$\rightarrow$$J=1$ $v=3$ line (with a velocity difference of about 1.5 km s$^{-1}$) acts on the population distribution of SiO. Thus, if both the H$_2$O $11_{6,6}$~$\nu_{2}=1$$\rightarrow$$12_{7,5}$~$\nu_{2}=0$ and $5_{0,5}$~$\nu_{2}=2$$\rightarrow$$6_{3,4}$~$\nu_{2}=1$ lines becomes stronger with increase of infrared colors, all correlations between infrared colors and the SiO maser intensity ratios among the $J=1$$-$0 $v=1$, 2 and 3 lines might be explained. The line intensity of the H$_2$O $5_{0,5}$~$\nu_{2}=2$$\rightarrow$$6_{3,4}$~$\nu_{2}=1$ line is usually weaker than that of the $11_{6,6}$~$\nu_{2}=1$$\rightarrow$$12_{7,5}$~$\nu_{2}=0$ line. This fact also seems to be consistent with the relatively weak intensity of the SiO $J=1$$-$0 $v=3$ line. In addition, as stated in Sect 3.2, the intensity of the $J=2$$-$1 $v=1$ line suddenly decreases at $\log(F_{25}/F_{12})\sim0.2$. If we rely on the population transfer by the H$_2$O overlap line, the behavior of the $J=2$$-$1 $v=1$ line seem to be explained consistently as suggested by \citet{nym93}: the overlapping H$_2$O line may become stronger and will therefore still be able to invert the 43 GHz transitions while the 86 GHz $v=1$ maser would become progressively weaker.

However, there are a couple of problems on the explanation with the overlap line of H$_2$O. First, we have to explain how the H$_2$O infrared lines overlapping with the SiO lines become stronger with increase of infrared colors. The relative abundance of H$_2$O molecules possibly increases with infrared colors, but this is not conclusive. Second, the correlation between infrared colors and the intensity ratios of the SiO $J=1$$-$0 $v=2$ to $v=1$ lines might be explained by more fundamental way without the overlap line of H$_2$O. In thick dusty envelopes of very red objects (i.e., envelopes exhibiting a large optical depth), strong 8~$\mu$m emission comes from every direction to the SiO masing region, causing ineffective pumping through the SiO $\Delta v=1$ transition, while 4~$\mu$m emission corresponding to the SiO $\Delta v=2$ transition is more effectively pump the SiO population instead of the 8~$\mu$m in such thick dusty envelopes (note that the 4~$\mu$m radiation only serves to pump the $v=2$ masers). Thus, the intensity ratio could be changed by the optical depth, which is uniquely determined by infrared colors such as $\log(F_{25}/ F_{12})$ and $\log(F_{\rm e}/ F_{\rm c})$. We have to also pay attention to the surface temperature of the central star, because the surface temperature of the central star affects on the 8~$\mu$m flux at the masing region, and also because the surface temperature correlates to the dust-color temperature.

In addition, we need to be careful about the reliability of the correlations seen in Figure 3, because the correlation between the MSX color [i.e., $\log(F_{\rm e}/ F_{\rm c})$] and the intensity ratios are not clear (see, panel E and F in Figure 3) comparing with the correlation between the IRAS color [i.e., $\log(F_{25}/ F_{12})$] and the SiO maser intensity ratios (see, panel B and C in Figure 3). One possible reason for the weak correlation between the MSX color and the intensity ratio is the lack of the detections in the blue MSX color region. The blue sources in the sample were selected from bright (nearby) objects to reduce the observing time, and these nearby objects are often popped out from the MSX survey region (i.e., Galactic disk: $|b|<6^{\circ}$). Another important factor causing the weak correlation coefficients is the scatter of the data points in Figure 3 due mainly to the pulsation of AGB stars. To reduce the effects of the pulsation, we have to measure the SiO maser line intensity at a certain pulsation phase (e.g., intensity maximum), and to do that we have to monitor the objects in infrared wavelengths or SiO maser lines; this would be a kind of big projects. Otherwise, future large scale monitoring projects like the Large Synoptic Survey Telescope \citep[LSST, ][]{tys02} might be helpful if the date is promptly released to the public after the observation.


\section{Summary}
In this research we observed 75 known SiO maser sources quasi-simultaneously in the SiO $J=1$$-$0, $v=0$, 1, 2, 3 and 4 lines, SiO $J=2$$-$1 $v=1$ and 2, $^{29}$SiO $J=1$$-$0 $v=0$ and $J=2$$-$1 $v=0$, and $^{30}$SiO $J=1$$-$0 $v=0$ lines. We also observed the targets in the H$_2$O 6$_{1,6}$$-$5$_{2,3}$ line under rainy/heavy cloudy condition. The sample continuously covers a very wide dust-temperature range from 150 K to 2000 K. The cold part of the sample (less than 250 K) was selected from cold SiO maser sources found in our recent surveys. The main results of the present observations are summarized as follows:

\begin{enumerate}
\item The correlation between infrared colors and the intensity ratio of the SiO $J=1$$-$0 $v=2$ to $v=1$ lines is confirmed as reported by \citet{nak03b}.

\item The intensity rations of SiO $J=1$$-$0 $v=3$ to $v=1$\&2 lines possibly correlate with infrared colors. In particular these ratios exhibit a relatively large correlation coefficient with the $\log(F_{25}/F_{12})$ color (IRAS color).
 
\item Two overlap line of H$_2$O (i.e., $11_{6,6}$~$\nu_{2}=1$$\rightarrow$$12_{7,5}$~$\nu_{2}=0$ and $5_{0,5}$~$\nu_{2}=2$$\rightarrow$$6_{3,4}$~$\nu_{2}=1$) might explain the correlations between the infrared colors and the SiO maser intensity ratios among the $J=1$$-$0 $v=1$, 2 and 3 lines, although the phenomena also might be explained by more fundamental ways if we take into account the variation of opacity from object to object.
\end{enumerate}

One important future work on the basis of the present observation would be VLBI observations in the SiO $J=1$$-$0 $v=3$ line, because we found several new sources bright enough for VLBI observations in the $J=1$$-$0 $v=3$ line. Comparison of the spatial distributions between the $J=1$$-$0 $v=3$ and other lines will be helpful to consider the SiO maser pump mechanism.


\acknowledgments
The present research has been supported by the Academia Sinica Institute of Astronomy \& Astrophysics and by the Laboratory for Astronomical Imaging at the University of Illinois. JN thanks Paul Ho for his constant encouragement. This research has made use of the SIMBAD and ADS databases.

\clearpage


\begin{figure}
\epsscale{.60}
\plotone{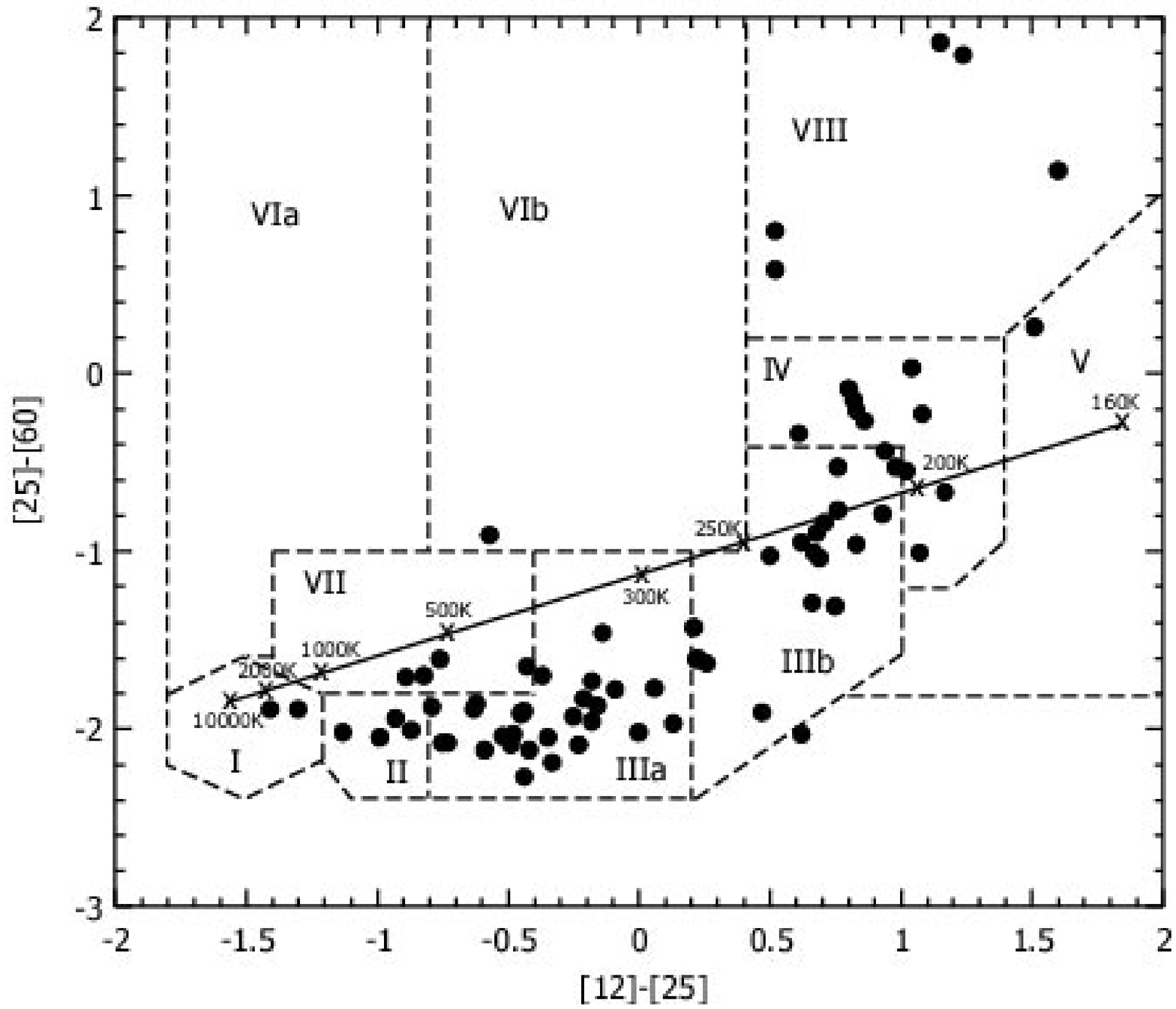}
\caption{Distribution of the observing targets in the IRAS two-color diagram. $[12]-[25]$ and $[25]-[60]$ mean the IRAS colors defined by equations as follows: $[12]-[25]=2.5\,\log(F_{\rm 25 \mu m}/F_{\rm 12 \mu m})$ and $[25]-[60]=2.5\,\log(F_{\rm 60 \mu m}/F_{\rm 25 \mu m})$, where $F_{\rm 12 \mu m}$, $F_{\rm 25 \mu m}$ and $F_{\rm 60 \mu m}$ are IRAS flux densities at the wavelength indicated in each suffix. The black dots represent the observing targets. The solid line represents the blackbody curve. The regions comparted by the dashed lines represents the classification of the IRAS sources suggested by \citet{van88}. \label{fig1}}
\end{figure}
\clearpage

\begin{figure}
\epsscale{1.0}
\plotone{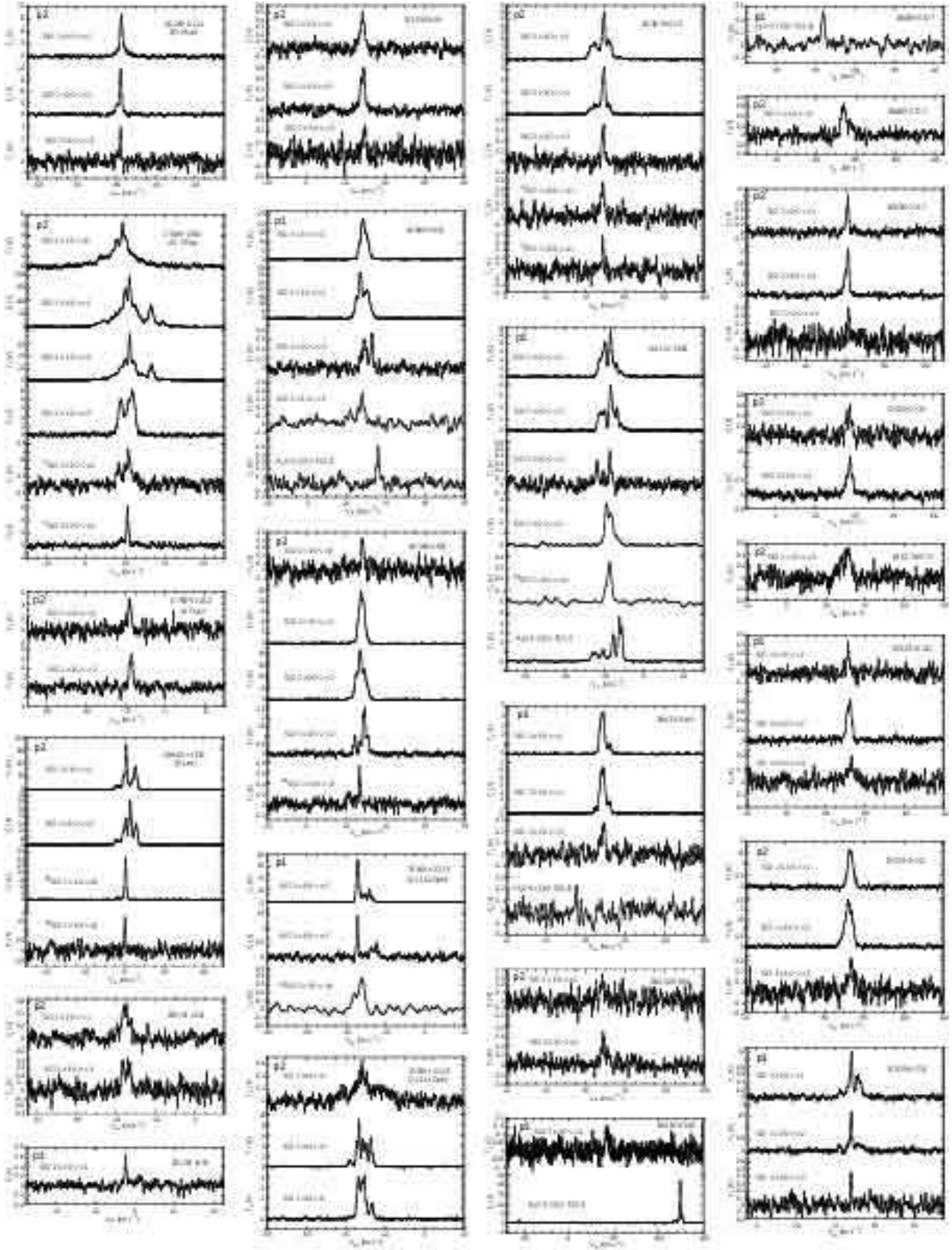}
\figcaption{($a$) Spectra of all detected lines. The designation ``p1'' and ``p2'' at the upper-left corners of each panel means the observing periods: May 11--19, 2004 (p1) and February 15--19, 2006 (p2). \label{fig2a}}
\end{figure}
\clearpage

\setcounter{figure}{1}
\begin{figure}
\epsscale{1.0}
\plotone{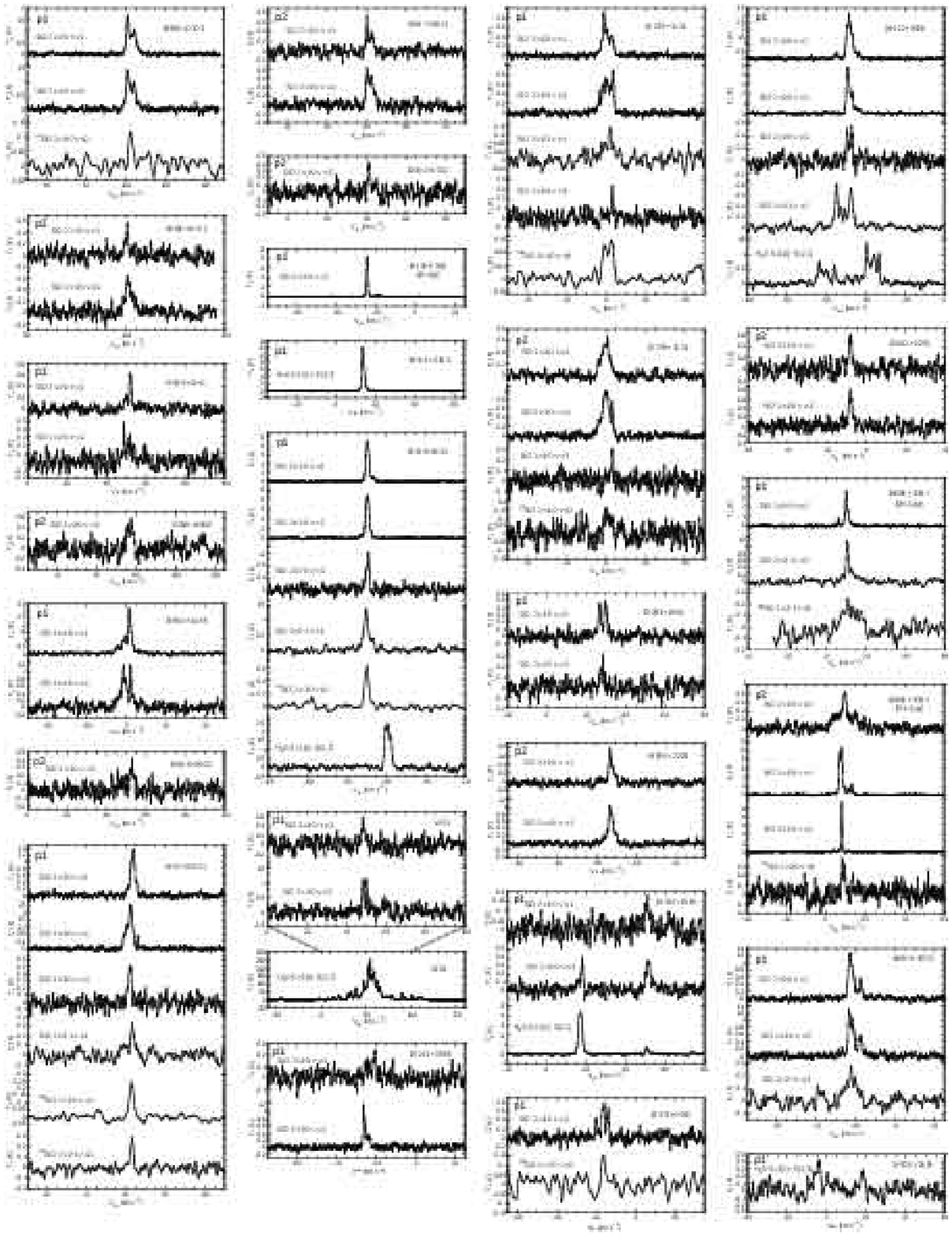}
\figcaption{($b$) Continued. \label{fig2b}}
\end{figure}
\clearpage

\setcounter{figure}{1}
\begin{figure}
\epsscale{1.0}
\plotone{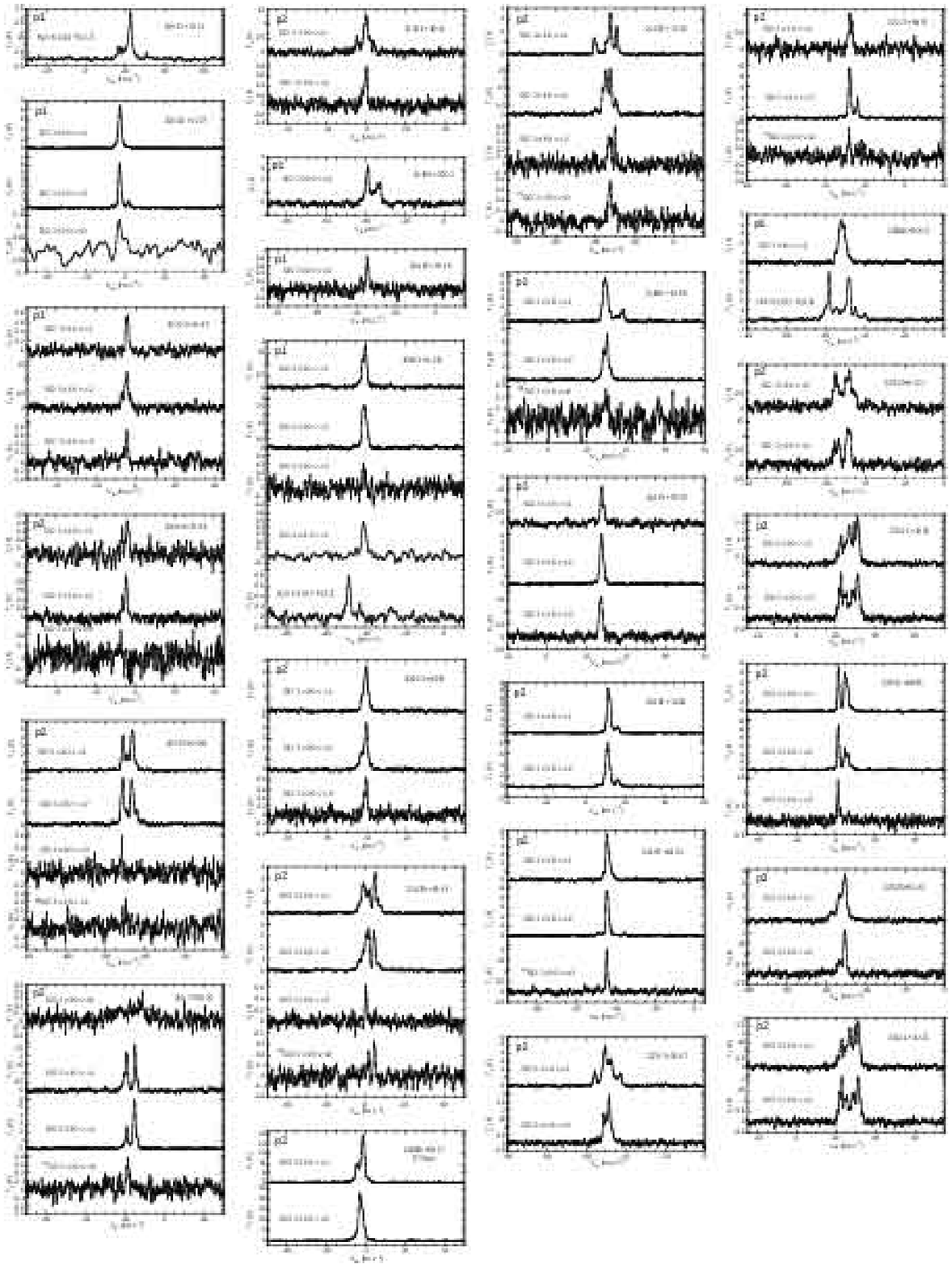}
\figcaption{($c$) Continued. \label{fig2c}}
\end{figure}
\clearpage

\begin{figure}
\epsscale{1.0}
\plotone{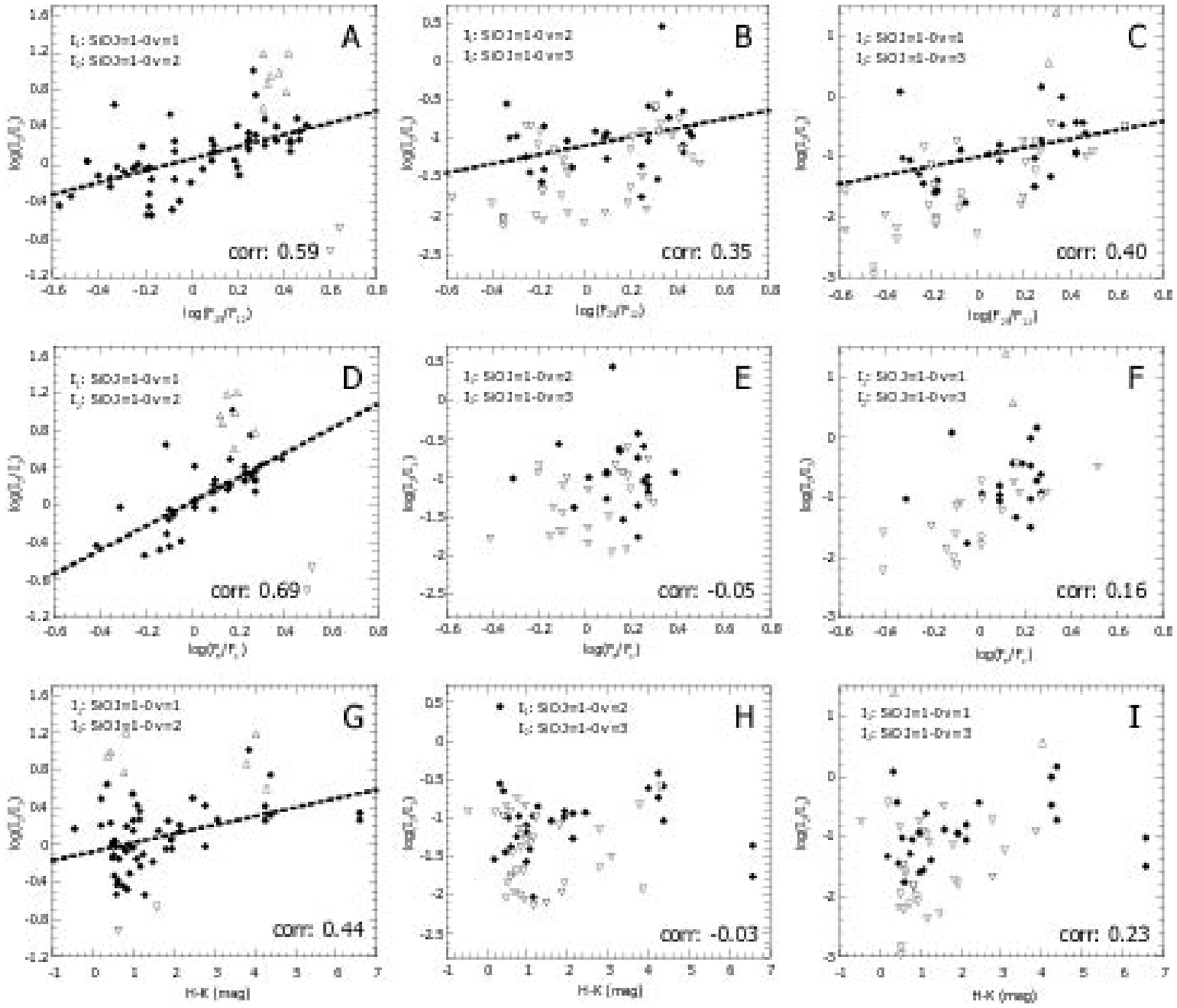}
\figcaption{Infrared colors versus intensity ratios of the SiO maser lines. The horizontal axes represent infrared colors. $F_{25}$ and $F_{12}$ denote the IRAS flux densities at $\lambda=25$ and 12~$\mu$m, respectively. $F_{\rm e}$ and $F_{\rm c}$ denote the MSX flux densities at $\lambda=21.3$ and 12.13~$\mu$m, respectively. The $H$$-$$K$ color was calculated from the 2MASS $H$- and $K$-bands magnitudes. $I_1$ and $I_2$ denote the velocity integrated intensities of each SiO maser line, and the assignment of the SiO lines to the $I_1$ and $I_2$ is indicated in each panel. The filled dots ($\bullet$), upward triangles ($\bigtriangleup$) and downward triangles ($\bigtriangledown$) respectively represent the intensity ratios of the SiO maser lines, lower limits of the ratio and upper limits of the ratio. Correlation coefficients are given in the lower-right corners of each panel. The dashed lines are the results of least-square fitting of a first order polynomial (only for the cases with the correlation coefficients larger than 0.35). \label{fig3}}
\end{figure}
\clearpage

\begin{figure}
\epsscale{1.0}
\plotone{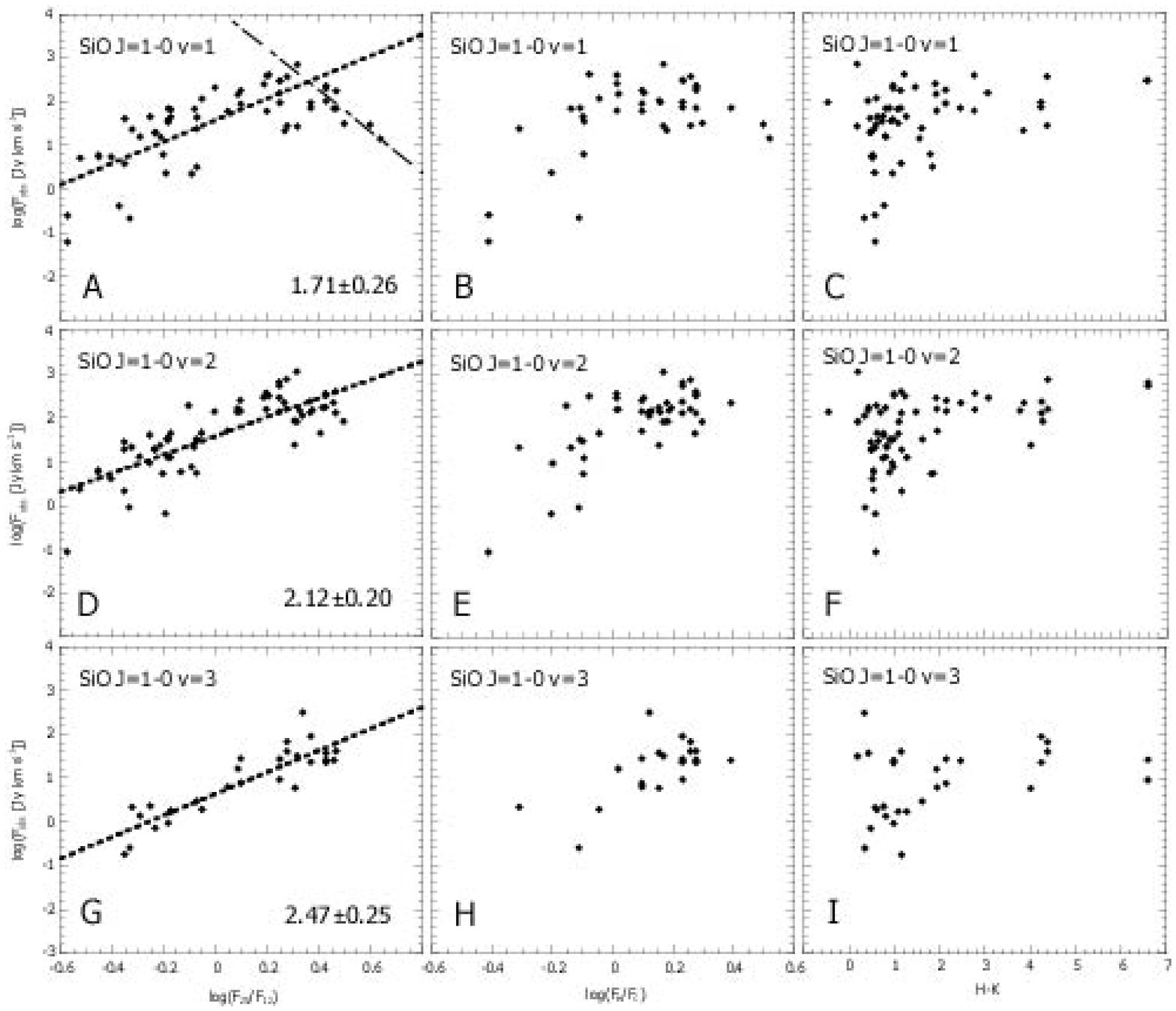}
\figcaption{Relation between infrared colors and absolute intensity of SiO maser lines. The notation of the infrared colors is same with that used in Figure 3. The intensity of SiO maser lines is standardized at the distance of 1 kpc using the luminosity distances given in Table 1. The thick dashed lines represent the results of least-square-fitting of a first order polynomial. The inclinations of the fitted lines (thick dashed lines) are given at the lower-right corners of each panel with statistical uncertainty. In the panel A, only the data points below $\log(F_{25}/F_{12})=0.5$ were fitted by the polynomial. The data points above $\log(F_{25}/F_{12})=0.5$ are independently fitted by a first order polynomial, and the results of the fitting is given as the thin chain line. \label{fig4}}
\end{figure}
\clearpage

\begin{figure}
\epsscale{1.0}
\plotone{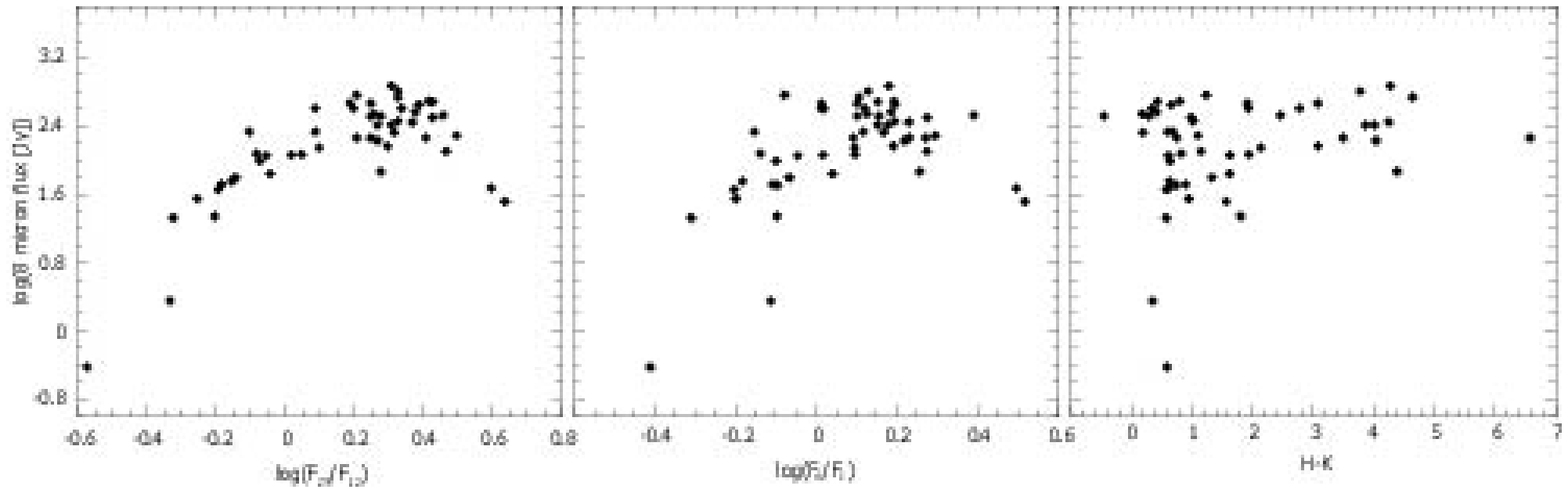}
\figcaption{Relations between 8~$\mu$m absolute flux and infrared colors. The 8~$\mu$m absolute flux is standardized at the distance of 1 kpc using the luminosity distances given in Table 1. \label{fig5}}
\end{figure}
\clearpage

\begin{figure}
\epsscale{0.45}
\plotone{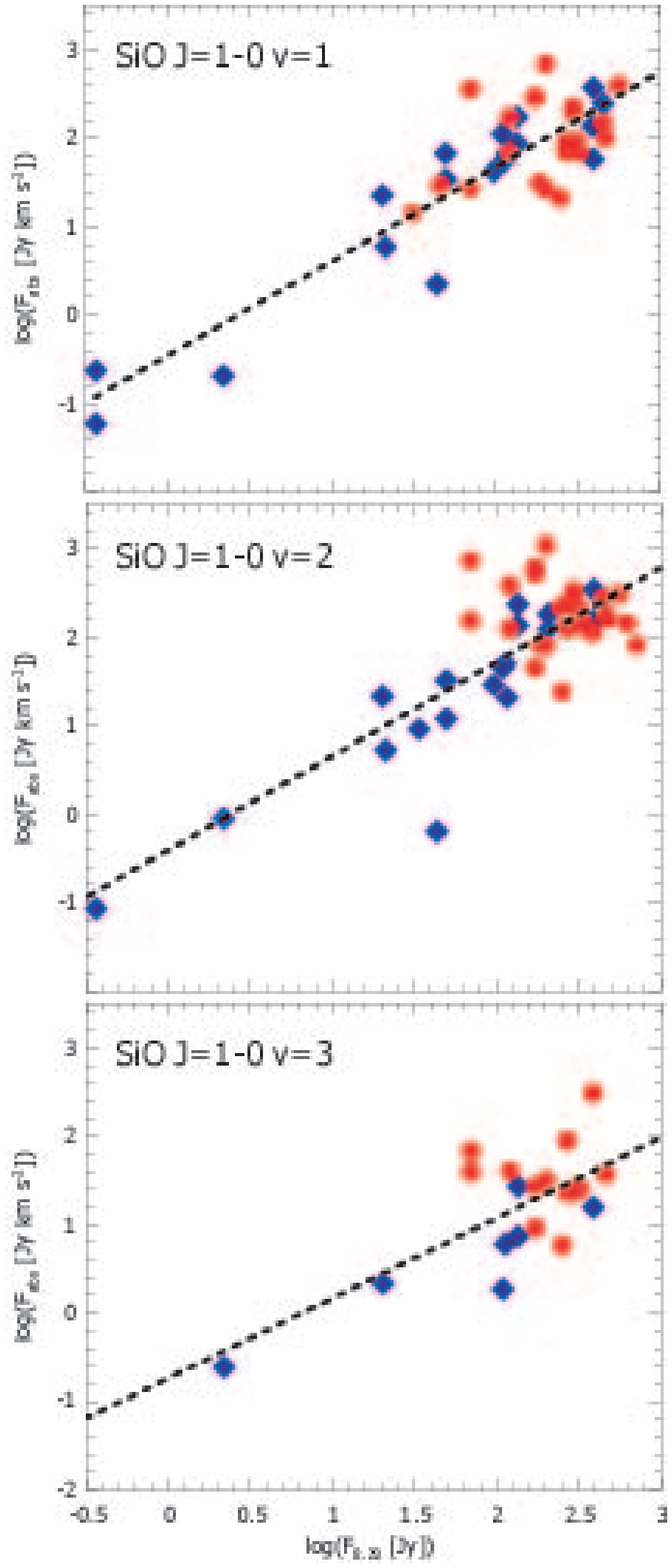}
\figcaption{Relation between 8~$\mu$m absolute flux and absolute intensities of the SiO $J=1$--0 $v=1$, 2 and 3 lines. The 8~$\mu$m absolute flux and the absolute intensities of the SiO maser lines are standardized at the distance of 1~kpc using the distances given in Table 1. The red circles and blue diamonds represent the objects exhibiting the $\log(F_{25}/F_{12})$ color above and below 0.2, respectively (i.e., the red circles represent relatively cold objects, whereas the blue diamonds represent relatively warm objects). The dashed lines represent the results of least-square-fitting of a first order polynomial. \label{fig6}}
\end{figure}
\clearpage

\begin{figure}
\epsscale{0.7}
\plotone{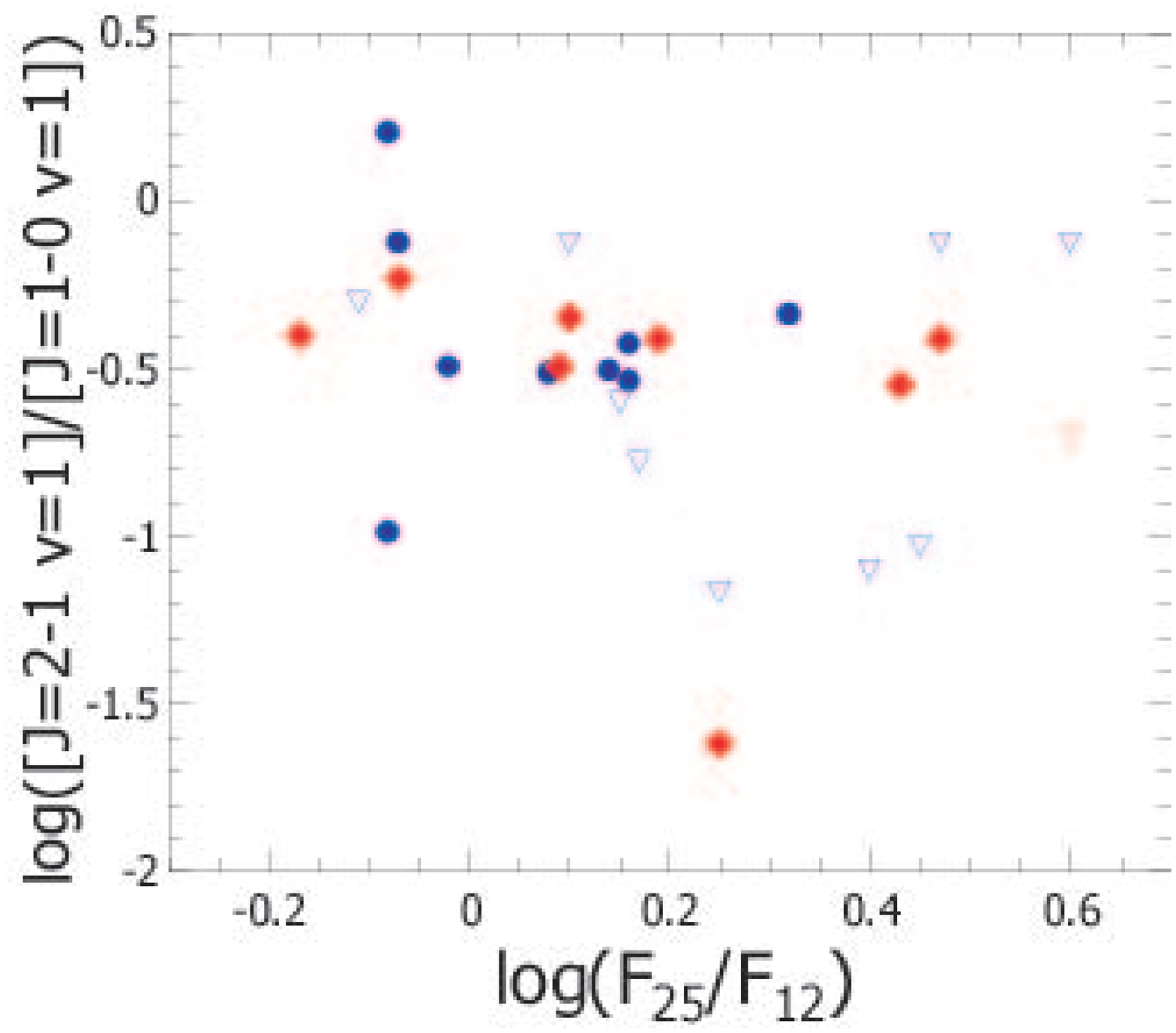}
\figcaption{$\log(F_{25}/F_{12})$ versus SiO maser intensity ratio of the $J=2$$-$1 $v=1$ to $J=1$$-$0 $v=1$ lines. As well as previous figures, $F_{25}$ and $F_{12}$ denote the IRAS flux densities at $\lambda=25$ and 12~$\mu$m, respectively. The red filled diamonds represent the results of the present observation. The blue filled dots represents the results of \citet{nym93}. The red and blue downward triangles respectively represent the upper limits in the present and Nyman's observations. \label{fig7}}
\end{figure}
\clearpage

\clearpage

\begin{deluxetable}{lrrrrrrrrr}
\tablecolumns{10}
\tablewidth{0pc}
\tablecaption{List of the observing targets}
\tablehead{
\colhead{IRAS name} &  \colhead{$J$} &  \colhead{$H$}  & \colhead{$K$} & \colhead{$F_{12}$} & \colhead{$F_{25}$} & \colhead{$F_{\rm a}$} & \colhead{$F_{\rm c}$} & \colhead{$F_{\rm e}$} & \colhead{$D$} \\
\colhead{} & \colhead{(mag)}   & \colhead{(mag)} & \colhead{(mag)} & \colhead{(Jy)} & \colhead{(Jy)} & \colhead{(Jy)} & \colhead{(Jy)} & \colhead{(Jy)} & \colhead{(kpc)}}
\startdata
06349$-$0121 & 2.3  & 1.4  & 0.8  & 148.1  & 71.2  & 96.0  & 102.2  & 50.3  & 0.46  \\
07209$-$2540 & 2.8  & 1.6  & 0.3  & 9919.0  & 6651.0  &  &  &  & ---  \\
07585$-$1242 & 2.4  & 1.6  & 1.1  & 91.3  & 53.2  &  &  &  & 0.79  \\
09448+1139 & $-$0.7  & $-$1.8  & $-$2.3  & 2161.0  & 654.0  &  &  &  & 0.06  \\
18035$-$2114 & 14.8  & 14.0  & 12.8  & 6.0  & 9.6  & 8.5  & 11.3  & 9.5  & 8.14  \\
18080$-$2238 & 11.7  & 10.9  & 10.7  & 11.1  & 20.4  & 9.3  & 14.1  & 18.9  & 6.05  \\
18135$-$1456 & 15.4  & 13.8  & 13.1  & 31.0  & 124.4  & 4.6  & 26.1  & 82.1  & 3.16  \\
18152$-$0919 & 14.6  & 13.6  & 13.2  & 14.5  & 38.8  & 17.7  & 24.6  & 35.1  & 5.11  \\
18199$-$1442 & 16.9  & 15.8  & 11.1  & 12.1  & 25.5  & 15.7  & 22.3  & 28.4  & 5.80  \\
18231$-$1112 & 16.9  & 13.4  & 9.9  & 10.6  & 17.2  & 4.6  & 7.0  & 8.7  & 6.11  \\
18242$-$0823 & 12.2  & 11.3  & 11.0  & 10.7  & 20.1  & 8.4  & 12.4  & 15.8  & 6.19  \\
18268$-$1117 & 13.8  & 12.3  & 11.7  & 12.4  & 30.5  & 13.9  & 20.2  & 31.9  & 5.62  \\
18348$-$0526 & 17.1  & 14.6  & 8.0  & 359.8  & 633.8  & 155.1  & 261.4  & 445.5  & 1.06  \\
18349+1023 & 4.1  & 2.3  & 1.1  & 719.7  & 318.5  &  &  &  & 0.18  \\
18387$-$0423 & 3.9  & 2.7  & 2.1  & 319.4  & 282.0  & 198.3  & 263.7  & 238.2  & 0.74  \\
18413+1354 & 4.7  & 3.2  & 2.1  & 225.1  & 152.3  &  &  &  & 0.62  \\
18432$-$0149 & 15.0  & 14.6  & 14.3  & 25.1  & 52.3  & 12.5  & 19.8  & 29.2  & 4.02  \\
18450$-$0148 & 16.1  & 14.8  & 13.2  & 23.7  & 103.5  & 2.5  & 23.9  & 78.8  & 3.53  \\
18488$-$0107 & 16.3  & 13.9  & 13.1  & 16.5  & 43.0  & 20.7  & 31.1  & 48.9  & 4.82  \\
18498$-$0017 & 16.8  & 16.0  & 13.5  & 22.9  & 65.8  & 20.2  & 29.7  & 73.4  & 4.01  \\
18509$-$0018 & 14.6  & 13.5  & 13.1  & 15.7  & 34.6  & 15.3  & 24.8  & 32.9  & 5.07  \\
18517+0037 & 16.7  & 15.7  & 15.3  & 16.6  & 39.4  & 14.8  & 25.9  & 39.8  & 4.88  \\
18525+0210 & 18.4  & 16.9  & 12.5  & 18.4  & 35.3  & 3.2  & 4.9  & 8.8  & 4.71  \\
18535+0726 & 13.1  & 10.2  & 8.2  & 63.0  & 70.7  & 25.5  & 30.4  & 38.0  & 2.11  \\
18540+0302 & 14.1  & 9.8  & 7.0  & 18.6  & 29.5  & 18.5  & 24.2  & 25.0  & 4.60  \\
18545+1040 & 10.6  & 8.4  & 6.6  & 48.2  & 30.6  & 13.9  & 18.0  & 14.5  & 1.23  \\
18549+0208 & 18.2  & 15.7  & 11.9  & 13.1  & 28.1  & 20.5  & 29.8  & 40.1  & 5.56  \\
18592+1455 & 6.4  & 4.9  & 4.0  & 37.7  & 25.2  & 22.1  & 23.6  & 18.5  & 1.50  \\
19017+0608 & 17.8  & 13.6  & 9.6  & 16.5  & 31.0  & 6.7  & 10.2  & 17.0  & 4.97  \\
19065+0832 & 9.3  & 7.6  & 6.9  & 20.2  & 51.5  & 9.1  & 16.4  & 30.8  & 4.37  \\
19067+0811 & 14.8  & 12.2  & 11.1  & 24.6  & 72.1  & 8.3  & 15.5  & 29.3  & 3.85  \\
19081+0322 & 16.7  & 15.0  & 10.7  & 9.1  & 18.4  & 16.1  & 21.7  & 33.1  & 6.69  \\
19126$-$0708 & 1.5  & 0.2  & $-$0.6  & 1575.0  & 669.6  &  &  &  & 0.12  \\
19128+0910 & 15.9  & 14.7  & 11.6  & 8.4  & 16.8  & 2.9  & 4.5  & 7.0  & 6.96  \\
19161+2343 & 13.4  & 9.6  & 6.9  & 112.2  & 137.2  &  &  &  & 1.68  \\
19192+0922 & 9.4  & 6.8  & 4.8  & 127.2  & 155.0  & 157.4  & 206.6  & 216.0  & 1.57  \\
19231+3555 & 6.0  & 4.6  & 3.6  & 111.9  & 90.4  &  &  &  & 1.12  \\
19252+2201 & 7.6  & 5.7  & 4.3  & 65.2  & 47.0  & 37.3  & 39.0  & 33.7  & 1.27  \\
19254+1631 & 17.8  & 15.8  & 14.8  & 16.6  & 44.7  & 13.1  & 22.1  & 41.7  & 4.77  \\
19283+1944 & 17.1  & 16.2  & 12.2  & 88.5  & 178.8  & 55.0  & 76.7  & 108.9  & 2.14  \\
19295+2228 & 15.2  & 12.6  & 9.5  & 14.3  & 25.3  & 15.8  & 22.8  & 28.9  & 5.34  \\
19312+1950 & 11.3  & 7.7  & 6.6  & 22.5  & 70.6  & 11.8  & 22.7  & 45.0  & 3.96  \\
19354+5005 & 2.3  & 1.4  & 0.9  & 104.9  & 52.2  &  &  &  & 0.58  \\
19374+0550 & 4.8  & 3.2  & 2.3  & 155.2  & 114.6  &  &  &  & 0.85  \\
19422+3506 & 9.2  & 6.9  & 5.3  & 202.9  & 172.3  &  &  &  & 0.89  \\
19440+2251 & 16.0  & 15.1  & 15.6  & 15.7  & 27.9  & 12.2  & 17.4  & 25.0  & 5.08  \\
19486+3247 & 0.2  & $-$1.1  & $-$1.7  & 1688.0  & 459.0  & 121.8  & 1748.2  & 680.1  & 0.05  \\
19493+2905 & 10.2  & 7.5  & 5.6  & 22.3  & 34.4  & 25.9  & 34.9  & 36.0  & 4.16  \\
19508+2659 & 9.6  & 6.9  & 5.3  & 28.1  & 29.7  & 12.5  & 17.9  & 18.6  & 3.00  \\
19576+2814 & 15.7  & 14.8  & 13.8  & 17.4  & 37.3  & 11.8  & 15.9  & 25.1  & 4.82  \\
20010+3011 & 5.0  & 3.7  & 3.1  & 58.2  & 41.4  & 31.7  & 50.2  & 33.1  & 1.32  \\
20024+1727 & 8.4  & 6.5  & 5.0  & 61.0  & 60.9  &  &  &  & 1.93  \\
20043+2653 & 17.4  & 14.9  & 10.6  & 17.9  & 42.0  & 12.3  & 17.2  & 29.4  & 4.70  \\
20052+0554 & 5.4  & 4.1  & 3.1  & 137.4  & 90.5  &  &  &  & 0.77  \\
20077$-$0625 & 6.9  & 3.9  & 2.1  & 1255.0  & 1061.0  &  &  &  & 0.36  \\
20095+2726 & 9.8  & 7.2  & 5.6  & 28.8  & 26.6  & 10.2  & 13.3  & 14.7  & 2.59  \\
20241+3811 & 2.4  & 1.0  & 0.4  & 510.6  & 329.6  & 289.8  & 482.6  & 303.6  & 0.39  \\
20381+5001 & 5.3  & 4.3  & 3.7  & 32.5  & 25.8  & 49.8  & 52.0  & 36.6  & 2.04  \\
20459+5015 & 6.6  & 5.1  & 4.2  & 20.3  & 11.4  & 13.7  & 13.7  & 8.7  & 1.57  \\
20491+4236 & 11.3  & 8.0  & 5.9  & 54.8  & 69.8  & 22.3  & 31.2  & 39.0  & 2.46  \\
20529+3013 & 2.9  & 1.9  & 1.4  & 171.6  & 101.4  &  &  &  & 0.59  \\
21088+6817 & $-$0.5  & $-$1.3  & $-$1.8  & 752.9  & 266.6  &  &  &  & 0.13  \\
21270+7135 & 3.8  & 2.7  & 1.9  & 105.7  & 54.2  &  &  &  & 0.60  \\
21286+1055 & 4.6  & 3.1  & 2.2  & 160.7  & 106.6  &  &  &  & 0.72  \\
21419+5832 & $-$0.3  & $-$1.3  & $-$1.6  & 1296.0  & 607.7  & 97.6  & 379.4  & 294.5  & 0.15  \\
21426+1228 & 2.7  & 1.8  & 1.3  & 59.5  & 26.7  &  &  &  & 0.65  \\
21439$-$0226 & $-$0.3  & $-$1.4  & $-$1.7  & 637.4  & 320.7  &  &  &  & 0.24  \\
21456+6422 & 4.0  & 2.7  & 1.9  & 175.3  & 108.1  &  &  &  & 0.62  \\
22097+5647 & 3.6  & 2.4  & 1.7  & 203.3  & 135.5  & 119.7  & 156.7  & 126.7  & 0.64  \\
22177+5936 & 16.4  & 10.7  & 6.9  & 123.2  & 228.9  & 75.2  & 103.0  & 155.8  & 1.82  \\
22480+6002 & 4.6  & 3.5  & 2.8  & 141.9  & 175.2  & 92.2  & 123.1  & 161.5  & 1.50  \\
22512+6100 & 3.9  & 2.6  & 2.0  & 108.5  & 93.3  & 63.5  & 83.7  & 67.0  & 1.23  \\
22516+0838 & 4.4  & 3.1  & 2.4  & 113.2  & 63.9  &  &  &  & 0.67  \\
22525+6033 & 4.6  & 3.0  & 2.1  & 112.1  & 92.3  & 87.8  & 133.6  & 97.7  & 1.15  \\
23041+1016 & 1.8  & 0.9  & 0.4  & 182.9  & 73.2  &  &  &  & 0.31  \\
\enddata
\tablenotetext{-}{$J$, $H$ and $K$ represent near-infrared $J$, $H$ and $K$-bands magnitudes taken from the 2MASS point source catalog. $F_{\rm 12}$ and $F_{\rm 25}$ represent IRAS flux densities at $\lambda=$12 and 15 $\mu$m, respectively. $F_{\rm a}$, $F_{\rm c}$ and $F_{\rm e}$ represent MSX flux densities at $\lambda=$8.28, 12.13 and 21.3 $\mu$m, respectively.}
\end{deluxetable}

\clearpage

\begin{deluxetable}{lrrr}
\tablecolumns{4}
\tablewidth{0pc}
\tablecaption{Statistics of the line observations}
\tablehead{
\colhead{Transition} & \colhead{Detections} & \colhead{No. of observations} & \colhead{Detection rate (\%)}}
\startdata
SiO $J=1$--0 $v=0$ & 5 & 53 & 9.4  \\
SiO $J=1$--0 $v=1$ & 60 & 78 & 76.9 \\
SiO $J=1$--0 $v=2$ & 68 & 83 & 81.9 \\
SiO $J=1$--0 $v=3$ & 29 & 79 & 36.7 \\
SiO $J=1$--0 $v=4$ & 0 & 50 & 0.0 \\
SiO $J=2$--1 $v=1$ & 8 & 9 & 88.9 \\
SiO $J=2$--1 $v=2$ & 1 & 23 & 4.3 \\
$^{29}$SiO $J=1$--0 $v=0$ & 21 & 81 & 25.9 \\
$^{29}$SiO $J=2$--1 $v=0$ & 2 & 23 & 8.7 \\
$^{30}$SiO $J=1$--0 $v=0$ & 3 & 51 & 5.9 \\
H$_2$O 6$_{1,6}$$-$5$_{2,3}$ & 14 & 28 & 50.0 \\
\enddata
\end{deluxetable}

\clearpage

\begin{deluxetable}{lcrrrrr}
\tablecolumns{7}
\tablewidth{0pc}
\tablecaption{Results of line observations}
\tablehead{
\colhead{IRAS name} & \colhead{transition} & \colhead{$V_{\rm lsr}$} &  \colhead{$T_{\rm peak}$}  & \colhead{$S$} &  \colhead{rms} & \colhead{obs. date} \\
\colhead{} & \colhead{} & \colhead{(km s$^{-1}$)} & \colhead{(K)} & \colhead{(K km s$^{-1}$)} & \colhead{(K)} & \colhead{(yymmdd.d)} }
\startdata
06349$-$0121 & SiO $J=1$--0 $v=0$ &  &  & $<$0.72 & 0.369  & 060215.9 \\
(SY Mon) & SiO $J=1$--0 $v=1$ & $-$56.9  & 16.99  & 37.19  & 0.424  & 060215.9 \\
 & SiO $J=1$--0 $v=2$ & $-$58.4  & 19.28  & 35.28  & 0.952  & 060215.9 \\
 & SiO $J=1$--0 $v=3$ & $-$57.8  & 2.90  & 3.48  & 0.414  & 060215.9 \\
 & $^{29}$SiO $J=1$--0 $v=0$ &  &  & $<$0.68 & 0.347  & 060215.9 \\
07209$-$2540 & SiO $J=1$--0 $v=0$ & 16.2  & 8.43  & 79.87  & 0.252  & 060216.0 \\
(VY CMa) & SiO $J=1$--0 $v=1$ & 24.8  & 193.03  & 1731.78  & 0.360  & 060216.0 \\
 & SiO $J=1$--0 $v=2$ & 22.8  & 77.81  & 492.98  & 0.280  & 060216.0 \\
 & SiO $J=1$--0 $v=3$ & 21.3  & 10.23  & 68.61  & 0.277  & 060216.0 \\
 & SiO $J=1$--0 $v=4$ &  &  & $<$0.51 & 0.258  & 060216.0 \\
 & $^{29}$SiO $J=1$--0 $v=0$ & 20.5  & 2.14  & 9.91  & 0.239  & 060216.0 \\
 & $^{30}$SiO $J=1$--0 $v=0$ & 19.7  & 6.21  & 10.06  & 0.252  & 060216.0 \\
07585$-$1242   & SiO $J=1$--0 $v=0$ &  &  & $<$1.28 & 0.656  & 060215.9 \\
(U Pup) & SiO $J=1$--0 $v=1$ & $-$18.3  & 4.91  & 10.28  & 0.560  & 060215.9 \\
 & SiO $J=1$--0 $v=2$ & $-$18.4  & 4.91  & 10.28  & 0.775  & 060215.9 \\
 & SiO $J=1$--0 $v=3$ &  &  & $<$1.48 & 0.748  & 060215.9 \\
 & $^{29}$SiO $J=1$--0 $v=0$ &  &  & $<$1.27 & 0.648  & 060215.9 \\
09448+1139  & SiO $J=1$--0 $v=0$ &  &  & $<$0.44 & 0.228  & 060216.0 \\
(R Leo) & SiO $J=1$--0 $v=1$ & 2.0  & 171.74  & 519.14  & 0.249  & 060216.0 \\
 & SiO $J=1$--0 $v=2$ & 0.7  & 61.77  & 238.45  & 0.235  & 060216.0 \\
 & $^{29}$SiO $J=1$--0 $v=0$ & $-$1.1  & 30.18  & 35.24  & 0.204  & 060216.0 \\
 & $^{30}$SiO $J=1$--0 $v=0$ & 0.1  & 1.78  & 1.70  & 0.236  & 060216.0 \\
18035$-$2114 & SiO $J=1$--0 $v=0$ &  &  & $<$0.14 & 0.069  & 060215.5 \\
 & SiO $J=1$--0 $v=1$ & $-$34.5  & 0.58  & 2.04  & 0.081  & 060215.5 \\
 & SiO $J=1$--0 $v=2$ & $-$34.9  & 0.43  & 1.58  & 0.079  & 060215.5 \\
 & SiO $J=1$--0 $v=3$ &  &  & $<$0.16 & 0.083  & 060215.5 \\
 & SiO $J=1$--0 $v=4$ &  &  & $<$0.15 & 0.074  & 060215.5 \\
 & $^{29}$SiO $J=1$--0 $v=0$ &  &  & $<$0.13 & 0.066  & 060215.5 \\
 & $^{30}$SiO $J=1$--0 $v=0$ &  &  & $<$0.15 & 0.076  & 060215.5 \\
18080$-$2238 & SiO $J=1$--0 $v=0$ &  &  & $<$0.22 & 0.114  & 060215.5 \\
 & SiO $J=1$--0 $v=1$ &  &  & $<$0.23 & 0.119  & 060215.5 \\
 & SiO $J=1$--0 $v=2$ &  &  & $<$0.23 & 0.117  & 060215.5 \\
 & SiO $J=1$--0 $v=3$ &  &  & $<$0.25 & 0.127  & 060215.5 \\
 & SiO $J=1$--0 $v=4$ &  &  & $<$0.23 & 0.114  & 060215.5 \\
 & $^{29}$SiO $J=1$--0 $v=0$ &  &  & $<$0.20 & 0.102  & 060215.5 \\
 & $^{30}$SiO $J=1$--0 $v=0$ &  &  & $<$0.24 & 0.121  & 060215.5 \\
18135$-$1456 & SiO $J=1$--0 $v=0$ &  &  & $<$0.20 & 0.101  & 060219.4 \\
 & SiO $J=1$--0 $v=1$ & $-$4.7  & 0.64  & 1.00  & 0.060  & 040512.2 \\
 &  &  &  & $<$0.99 & 0.110  & 060219.4 \\
 & SiO $J=1$--0 $v=2$ &  &  & $<$0.12 & 0.062  & 040512.2 \\
 &  &  &  & $<$0.95 & 0.106  & 060219.4 \\
 & SiO $J=1$--0 $v=3$ &  &  & $<$0.24 & 0.121  & 060219.4 \\
 & SiO $J=1$--0 $v=4$ &  &  & $<$0.22 & 0.109  & 060219.4 \\
 & SiO $J=2$--1 $v=1$ &  &  & $<$0.14 & 0.042  & 040511.2 \\
 & SiO $J=2$--1 $v=2$ &  &  & $<$0.19 & 0.055  & 040512.2 \\
 & $^{29}$SiO $J=1$--0 $v=0$ &  &  & $<$0.11 & 0.023  & 040511.2 \\
 &  &  &  & $<$0.89 & 0.099  & 060219.4 \\
 & $^{29}$SiO $J=2$--0 $v=0$ &  &  & $<$0.19 & 0.055  & 040512.2 \\
 & $^{30}$SiO $J=1$--0 $v=0$ &  &  & $<$0.21 & 0.108  & 060219.4 \\
18152$-$0919 & SiO $J=1$--0 $v=0$ &  &  & $<$0.11 & 0.055  & 060215.5 \\
 & SiO $J=1$--0 $v=1$ & 28.5  & 0.51  & 1.31  & 0.059  & 060215.5 \\
 & SiO $J=1$--0 $v=2$ & 28.6  & 0.80  & 2.21  & 0.070  & 060215.5 \\
 & SiO $J=1$--0 $v=3$ & 29.4  & 0.26  & 0.48  & 0.059  & 060215.5 \\
 & SiO $J=1$--0 $v=4$ &  &  & $<$0.11 & 0.055  & 060215.5 \\
 & $^{29}$SiO $J=1$--0 $v=0$ &  &  & $<$0.10 & 0.051  & 060215.5 \\
 & $^{30}$SiO $J=1$--0 $v=0$ &  &  & $<$0.10 & 0.053  & 060215.5 \\
18199$-$1442 & SiO $J=1$--0 $v=0$ &  &  & $<$0.20 & 0.101  & 060219.4 \\
 & SiO $J=1$--0 $v=1$ &  &  & $<$0.21 & 0.108  & 060219.4 \\
 & SiO $J=1$--0 $v=2$ &  &  & $<$0.21 & 0.109  & 060219.4 \\
 & SiO $J=1$--0 $v=3$ &  &  & $<$0.24 & 0.119  & 060219.4 \\
 & SiO $J=1$--0 $v=4$ &  &  & $<$0.21 & 0.107  & 060219.4 \\
 & $^{29}$SiO $J=1$--0 $v=0$ &  &  & $<$0.19 & 0.095  & 060219.4 \\
 & $^{30}$SiO $J=1$--0 $v=0$ &  &  & $<$0.21 & 0.107  & 060219.4 \\
18231$-$1112 & SiO $J=1$--0 $v=0$ &  &  & $<$0.20 & 0.105  & 060215.5 \\
 & SiO $J=1$--0 $v=1$ &  &  & $<$0.21 & 0.106  & 060215.5 \\
 & SiO $J=1$--0 $v=2$ &  &  & $<$0.20 & 0.102  & 060215.5 \\
 & SiO $J=1$--0 $v=3$ &  &  & $<$0.22 & 0.112  & 060215.5 \\
 & SiO $J=1$--0 $v=4$ &  &  & $<$0.20 & 0.103  & 060215.5 \\
 & $^{29}$SiO $J=1$--0 $v=0$ &  &  & $<$0.18 & 0.094  & 060215.5 \\
 & $^{30}$SiO $J=1$--0 $v=0$ &  &  & $<$0.20 & 0.099  & 060215.5 \\
18242$-$0823 & SiO $J=1$--0 $v=0$ &  &  & $<$0.27 & 0.139  & 060215.5 \\
 & SiO $J=1$--0 $v=1$ &  &  & $<$0.27 & 0.135  & 060215.5 \\
 & SiO $J=1$--0 $v=2$ &  &  & $<$0.25 & 0.127  & 060215.5 \\
 & SiO $J=1$--0 $v=3$ &  &  & $<$0.28 & 0.142  & 060215.5 \\
 & SiO $J=1$--0 $v=4$ &  &  & $<$0.26 & 0.131  & 060215.5 \\
 & $^{29}$SiO $J=1$--0 $v=0$ &  &  & $<$0.23 & 0.119  & 060215.5 \\
 & $^{30}$SiO $J=1$--0 $v=0$ &  &  & $<$0.27 & 0.136  & 060215.5 \\
18268$-$1117 & SiO $J=1$--0 $v=0$ &  &  & $<$0.20 & 0.101  & 060219.4 \\
 & SiO $J=1$--0 $v=1$ &  &  & $<$0.21 & 0.106  & 060219.4 \\
 & SiO $J=1$--0 $v=2$ &  &  & $<$0.20 & 0.103  & 060219.4 \\
 & SiO $J=1$--0 $v=3$ &  &  & $<$0.22 & 0.114  & 060219.4 \\
 & SiO $J=1$--0 $v=4$ &  &  & $<$0.22 & 0.111  & 060219.4 \\
 & $^{29}$SiO $J=1$--0 $v=0$ &  &  & $<$0.19 & 0.097  & 060219.4 \\
 & $^{30}$SiO $J=1$--0 $v=0$ &  &  & $<$0.20 & 0.103  & 060219.4 \\
18348$-$0526 & SiO $J=1$--0 $v=0$ & 28.2  & 0.42  & 0.74  & 0.071  & 060215.5 \\
(V437 Sct) & SiO $J=1$--0 $v=1$ & 29.9  & 22.73  & 88.43  & 0.055  & 040511.3 \\
 &  & 27.3  & 24.50  & 86.91  & 0.077  & 060215.5 \\
 & SiO $J=1$--0 $v=2$ & 28.6  & 31.94  & 160.19  & 0.067  & 040511.3 \\
 &  & 26.8  & 42.91  & 187.78  & 0.079  & 060215.5 \\
 & SiO $J=1$--0 $v=3$ & 30.2  & 0.72  & 2.75  & 0.078  & 040515.2 \\
 &  & 27.5  & 2.58  & 8.08  & 0.086  & 060215.5 \\
 & SiO $J=1$--0 $v=4$ &  &  & $<$0.15 & 0.077  & 060215.5 \\
 & SiO $J=2$--1 $v=1$ & 25.7  & 0.37  & 1.52  & 0.043  & 040511.3 \\
 & SiO $J=2$--1 $v=2$ &  &  & $<$0.19 & 0.055  & 040512.3 \\
 & $^{29}$SiO $J=1$--0 $v=0$ & 24.1  & 0.73  & 1.29  & 0.066  & 060215.5 \\
 &  &  &  & $<$0.13 & 0.027  & 040512.2 \\
 & $^{30}$SiO $J=1$--0 $v=0$ &  &  & $<$0.14 & 0.072  & 060215.5 \\
 & H$_2$O 6$_{1,6}$$-$5$_{2,3}$ & 26.8  & 0.50  & 0.80  & 0.055  & 040510.2 \\
18349+1023 & SiO $J=1$--0 $v=0$ & $-$30.1 & 0.66  & 5.85  & 0.067  & 060215.5 \\
(V1111 Oph) & SiO $J=1$--0 $v=1$ & $-$32.0  & 8.98  & 37.61  & 0.069  & 060215.5 \\
 & SiO $J=1$--0 $v=2$ & $-$31.0  & 70.43  & 192.47  & 0.074  & 040515.2 \\
 &  & $-$31.7  & 4.17  & 21.89  & 0.072  & 060215.5 \\
 & SiO $J=1$--0 $v=3$ & $-$34.4  & 1.45  & 1.79  & 0.078  & 040515.2 \\
 &  &  &  & $<$0.16 & 0.081  & 060215.5 \\
 & SiO $J=1$--0 $v=4$ &  &  & $<$0.14 & 0.069  & 060215.5 \\
 & $^{29}$SiO $J=1$--0 $v=0$ & $-$32.0  & 0.38  & 1.78  & 0.087  & 040515.2 \\
 &  &  &  & $<$0.12 & 0.064  & 060215.5 \\
 & $^{30}$SiO $J=1$--0 $v=0$ &  &  & $<$0.14 & 0.069  & 060215.5 \\
18387$-$0423 & SiO $J=1$--0 $v=0$ &  &  & $<$0.15 & 0.076  & 060219.5 \\
 & SiO $J=1$--0 $v=1$ & 48.9  & 14.09  & 70.19  & 0.086  & 060219.5 \\
 & SiO $J=1$--0 $v=2$ & 46.8  & 8.28  & 28.50  & 0.077  & 060219.5 \\
 & SiO $J=1$--0 $v=3$ & 48.9  & 0.71  & 1.18  & 0.087  & 060219.5 \\
 & SiO $J=1$--0 $v=4$ &  &  & $<$0.15 & 0.077  & 060219.5 \\
 & $^{29}$SiO $J=1$--0 $v=0$ & 48.6  & 0.49  & 0.98  & 0.071  & 060219.5 \\
 & $^{30}$SiO $J=1$--0 $v=0$ & 48.8  & 0.49  & 0.55  & 0.081  & 060219.5 \\
18413+1354 & SiO $J=1$--0 $v=1$ & $-$16.8  & 11.37  & 56.20  & 0.095  & 040511.3 \\
(V837 Her) & SiO $J=1$--0 $v=2$ & $-$16.7  & 7.87  & 39.08  & 0.116  & 040511.3 \\
 & SiO $J=1$--0 $v=3$ & $-$20.5  & 0.48  & 1.53  & 0.053  & 040515.2 \\
 & SiO $J=2$--1 $v=1$ & $-$16.7  & 3.64  & 16.09  & 0.076  & 040511.3 \\
 & SiO $J=2$--1 $v=2$ &  &  & $<$0.12 & 0.036  & 040512.3 \\
 & $^{29}$SiO $J=1$--0 $v=0$ & $-$18.3  & 0.32  & 2.14  & 0.023  & 040515.2 \\
 & $^{29}$SiO $J=2$--0 $v=0$ &  &  & $<$0.12 & 0.036  & 040512.3 \\
 & H$_2$O 6$_{1,6}$$-$5$_{2,3}$ & $-$18.7  & 4.63  & 22.26  & 0.052  & 040510.2 \\
18432$-$0149 & SiO $J=1$--0 $v=0$ &  &  & $<$0.17 & 0.086  & 060218.5 \\
(V1360 Aql) & SiO $J=1$--0 $v=1$ & 69.2  & 4.34  & 14.45  & 0.047  & 040519.2 \\
 &  & 68.3  & 0.33  & 0.56  & 0.097  & 060218.5 \\
 & SiO $J=1$--0 $v=2$ & 68.6  & 6.44  & 23.08  & 0.044  & 040518.2 \\
 &  & 69.6  & 0.69  & 1.70  & 0.095  & 060218.5 \\
 & SiO $J=1$--0 $v=3$ & 68.7  & 0.27  & 0.67  & 0.047  & 040518.2 \\
 &  &  &  & $<$0.20 & 0.103  & 060218.5 \\
 & SiO $J=1$--0 $v=4$ &  &  & $<$0.20 & 0.099  & 060218.5 \\
 & SiO $J=2$--1 $v=2$ &  &  & $<$0.13 & 0.038  & 040519.2 \\
 & $^{29}$SiO $J=1$--0 $v=0$ &  &  & $<$0.12 & 0.025  & 040519.2 \\
 &  &  &  & $<$0.17 & 0.086  & 060218.5 \\
 & $^{29}$SiO $J=2$--0 $v=0$ &  &  & $<$0.13 & 0.037  & 040519.2 \\
 & $^{30}$SiO $J=1$--0 $v=0$ &  &  & $<$0.19 & 0.097  & 060218.5 \\
 & H$_2$O 6$_{1,6}$$-$5$_{2,3}$ & 55.2  & 0.23  & 0.28  & 0.062  & 040510.2 \\
18450$-$0148 & SiO $J=1$--0 $v=1$ & 41.4  & 0.16  & 0.38  & 0.037  & 040514.2 \\
(W43A) & SiO $J=1$--0 $v=2$ &  &  & $<$0.08 & 0.042  & 040514.2 \\
 & SiO $J=1$--0 $v=3$ &  &  & $<$0.13 & 0.063  & 040520.2 \\
 & SiO $J=2$--1 $v=2$ &  &  & $<$0.12 & 0.035  & 040514.2 \\
 & $^{29}$SiO $J=1$--0 $v=0$ &  &  & $<$0.08 & 0.017  & 040514.2 \\
 & $^{29}$SiO $J=2$--0 $v=0$ &  &  & $<$0.12 & 0.035  & 040514.2 \\
 & H$_2$O 6$_{1,6}$$-$5$_{2,3}$ & 35.2  & 45.38  & 103.62  & 0.126  & 040510.3 \\
18488$-$0107 & SiO $J=1$--0 $v=0$ &  &  & $<$0.15 & 0.079  & 060218.5 \\
(V1363 Aql) & SiO $J=1$--0 $v=1$ &  &  & $<$0.15 & 0.078  & 060218.5 \\
 & SiO $J=1$--0 $v=2$ & 75.4  & 0.66  & 2.43  & 0.080  & 060218.5 \\
 & SiO $J=1$--0 $v=3$ &  &  & $<$0.18 & 0.089  & 060218.5 \\
 & SiO $J=1$--0 $v=4$ &  &  & $<$0.16 & 0.083  & 060218.5 \\
 & $^{29}$SiO $J=1$--0 $v=0$ &  &  & $<$0.15 & 0.075  & 060218.5 \\
 & $^{30}$SiO $J=1$--0 $v=0$ &  &  & $<$0.16 & 0.083  & 060218.5 \\
 & H$_2$O 6$_{1,6}$$-$5$_{2,3}$ & 63.6  & 0.51  & 1.50  & 0.058  & 040517.2 \\
18498$-$0017 & SiO $J=1$--0 $v=0$ &  &  & $<$0.14 & 0.070  & 060215.5 \\
(V1365 Aql) & SiO $J=1$--0 $v=1$ & 61.1  & 1.03  & 1.46  & 0.085  & 060215.5 \\
 & SiO $J=1$--0 $v=2$ & 60.7  & 2.16  & 4.55  & 0.129  & 060215.5 \\
 & SiO $J=1$--0 $v=3$ & 61.9  & 0.39  & 0.53  & 0.082  & 060215.5 \\
 & SiO $J=1$--0 $v=4$ &  &  & $<$0.15 & 0.075  & 060215.5 \\
 & $^{29}$SiO $J=1$--0 $v=0$ &  &  & $<$0.12 & 0.064  & 060215.5 \\
 & $^{30}$SiO $J=1$--0 $v=0$ &  &  & $<$0.14 & 0.073  & 060215.5 \\
18509$-$0018 & SiO $J=1$--0 $v=0$ &  &  & $<$0.16 & 0.082  & 060218.5 \\
 & SiO $J=1$--0 $v=1$ &  &  & $<$0.16 & 0.082  & 060218.5 \\
 & SiO $J=1$--0 $v=2$ & 36.7  & 0.63  & 1.48  & 0.098  & 060218.5 \\
 & SiO $J=1$--0 $v=3$ & 37.1  & 1.36  & 4.04  & 0.105  & 060218.5 \\
 & SiO $J=1$--0 $v=4$ &  &  & $<$0.19 & 0.097  & 060218.5 \\
 & $^{29}$SiO $J=1$--0 $v=0$ &  &  & $<$0.16 & 0.082  & 060218.5 \\
 & $^{30}$SiO $J=1$--0 $v=0$ &  &  & $<$0.19 & 0.096  & 060218.5 \\
18517+0037 & SiO $J=1$--0 $v=0$ &  &  & $<$0.19 & 0.099  & 060218.5 \\
 & SiO $J=1$--0 $v=1$ &  &  & $<$0.21 & 0.108  & 060218.5 \\
 & SiO $J=1$--0 $v=2$ & 29.9  & 0.53  & 2.10  & 0.102  & 060218.5 \\
 & SiO $J=1$--0 $v=3$ &  &  & $<$0.22 & 0.113  & 060218.5 \\
 & SiO $J=1$--0 $v=4$ &  &  & $<$0.21 & 0.105  & 060218.5 \\
 & $^{29}$SiO $J=1$--0 $v=0$ &  &  & $<$0.18 & 0.092  & 060218.5 \\
 & $^{30}$SiO $J=1$--0 $v=0$ &  &  & $<$0.21 & 0.107  & 060218.5 \\
18525+0210 & SiO $J=1$--0 $v=0$ &  &  & $<$0.09 & 0.047  & 060218.5 \\
 & SiO $J=1$--0 $v=1$ & 71.2  & 0.32  & 0.43  & 0.046  & 040519.2 \\
 &  & 73.2  & 1.64  & 5.53  & 0.056  & 060218.5 \\
 & SiO $J=1$--0 $v=2$ & 71.7  & 0.84  & 2.38  & 0.040  & 040518.2 \\
 &  & 71.5  & 2.54  & 11.39  & 0.058  & 060218.5 \\
 & SiO $J=1$--0 $v=3$ & 72.8  & 0.20  & 0.61  & 0.042  & 040518.2 \\
 &  & 74.0  & 0.33  & 1.03  & 0.057  & 060218.5 \\
 & SiO $J=1$--0 $v=4$ &  &  & $<$0.10 & 0.053  & 060218.5 \\
 & SiO $J=2$--1 $v=2$ &  &  & $<$0.14 & 0.039  & 040519.2 \\
 & $^{29}$SiO $J=1$--0 $v=0$ &  &  & $<$0.12 & 0.025  & 040519.2 \\
 &  &  &  & $<$0.10 & 0.050  & 060218.5 \\
 & $^{29}$SiO $J=2$--0 $v=0$ &  &  & $<$0.13 & 0.037  & 040519.2 \\
 & $^{30}$SiO $J=1$--0 $v=0$ &  &  & $<$0.11 & 0.053  & 060218.5 \\
 & H$_2$O 6$_{1,6}$$-$5$_{2,3}$ &  &  & $<$0.15 & 0.054  & 040517.2 \\
18535+0726 & SiO $J=1$--0 $v=1$ & 47.7  & 1.18  & 4.39  & 0.045  & 040519.3 \\
 & SiO $J=1$--0 $v=2$ & 49.7  & 1.69  & 3.92  & 0.069  & 040518.3 \\
 & SiO $J=1$--0 $v=3$ & 47.9  & 0.46  & 0.47  & 0.071  & 040518.3 \\
 & SiO $J=2$--1 $v=2$ &  &  & $<$0.12 & 0.035  & 040519.3 \\
 & $^{29}$SiO $J=1$--0 $v=0$ &  &  & $<$0.12 & 0.025  & 040519.3 \\
 & $^{29}$SiO $J=2$--0 $v=0$ &  &  & $<$0.12 & 0.036  & 040519.3 \\
 & H$_2$O 6$_{1,6}$$-$5$_{2,3}$ &  &  & $<$0.20 & 0.074  & 040517.3 \\
18540+0302 & SiO $J=1$--0 $v=0$ &  &  & $<$0.15 & 0.078  & 060218.5 \\
 & SiO $J=1$--0 $v=1$ & 103.0  & 1.68  & 6.09  & 0.051  & 040519.3 \\
 &  & 100.1  & 0.57  & 0.95  & 0.087  & 060218.5 \\
 & SiO $J=1$--0 $v=2$ & 102.9  & 1.44  & 5.76  & 0.066  & 040520.2 \\
 &  & 101.7  & 0.66  & 2.47  & 0.083  & 060218.5 \\
 & SiO $J=1$--0 $v=3$ &  &  & $<$0.13 & 0.066  & 040520.2 \\
 &  &  &  & $<$0.18 & 0.090  & 060218.5 \\
 & SiO $J=1$--0 $v=4$ &  &  & $<$0.17 & 0.087  & 060218.5 \\
 & SiO $J=2$--1 $v=2$ &  &  & $<$0.15 & 0.044  & 040519.3 \\
 & $^{29}$SiO $J=1$--0 $v=0$ & 102.1  & 0.12  & 0.30  & 0.028  & 040519.3 \\
 &  &  &  & $<$0.15 & 0.075  & 060218.5 \\
 & $^{29}$SiO $J=2$--0 $v=0$ &  &  & $<$0.15 & 0.043  & 040519.3 \\
 & $^{30}$SiO $J=1$--0 $v=0$ &  &  & $<$0.17 & 0.086  & 060218.5 \\
18545+1040 & SiO $J=1$--0 $v=1$ & 51.1  & 0.64  & 1.35  & 0.056  & 040519.3 \\
 & SiO $J=1$--0 $v=2$ & 50.4  & 0.48  & 1.20  & 0.072  & 040519.3 \\
 & SiO $J=1$--0 $v=3$ &  &  & $<$0.10 & 0.049  & 040518.3 \\
 & SiO $J=2$--1 $v=2$ &  &  & $<$0.16 & 0.046  & 040519.3 \\
 & $^{29}$SiO $J=1$--0 $v=0$ &  &  & $<$0.14 & 0.028  & 040519.3 \\
 & $^{29}$SiO $J=2$--0 $v=0$ &  &  & $<$0.16 & 0.046  & 040519.3 \\
 & H$_2$O 6$_{1,6}$$-$5$_{2,3}$ &  &  & $<$0.19 & 0.070  & 040517.3 \\
18549+0208 & SiO $J=1$--0 $v=0$ &  &  & $<$0.20 & 0.103  & 060218.5 \\
 & SiO $J=1$--0 $v=1$ &  &  & $<$0.21 & 0.109  & 060218.5 \\
 & SiO $J=1$--0 $v=2$ & 76.7  & 0.59  & 1.60  & 0.110  & 060218.5 \\
 & SiO $J=1$--0 $v=3$ &  &  & $<$0.24 & 0.122  & 060218.5 \\
 & SiO $J=1$--0 $v=4$ &  &  & $<$0.23 & 0.115  & 060218.5 \\
 & $^{29}$SiO $J=1$--0 $v=0$ &  &  & $<$0.20 & 0.101  & 060218.5 \\
 & $^{30}$SiO $J=1$--0 $v=0$ &  &  & $<$0.23 & 0.115  & 060218.5 \\
18592+1455 & SiO $J=1$--0 $v=1$ & 1.4  & 3.20  & 10.31  & 0.066  & 040519.3 \\
 & SiO $J=1$--0 $v=2$ & $-$0.5  & 1.19  & 4.98  & 0.092  & 040519.3 \\
 & SiO $J=1$--0 $v=3$ &  &  & $<$0.10 & 0.052  & 040518.3 \\
 & SiO $J=2$--1 $v=2$ &  &  & $<$0.19 & 0.054  & 040519.3 \\
 & $^{29}$SiO $J=1$--0 $v=0$ &  &  & $<$0.18 & 0.036  & 040519.3 \\
 & $^{29}$SiO $J=2$--0 $v=0$ &  &  & $<$0.18 & 0.054  & 040519.3 \\
19017+0608 & SiO $J=1$--0 $v=0$ &  &  & $<$0.19 & 0.096  & 060217.5 \\
(V1367 Aql) & SiO $J=1$--0 $v=1$ &  &  & $<$0.21 & 0.105  & 060217.5 \\
 & SiO $J=1$--0 $v=2$ &  &  & $<$0.20 & 0.104  & 060217.5 \\
 & SiO $J=1$--0 $v=3$ &  &  & $<$0.22 & 0.113  & 060217.5 \\
 & SiO $J=1$--0 $v=4$ &  &  & $<$0.21 & 0.106  & 060217.5 \\
 & $^{29}$SiO $J=1$--0 $v=0$ &  &  & $<$0.18 & 0.091  & 060217.5 \\
 & $^{30}$SiO $J=1$--0 $v=0$ &  &  & $<$0.21 & 0.105  & 060217.5 \\
19065+0832 & SiO $J=1$--0 $v=0$ &  &  & $<$0.14 & 0.070  & 060215.5 \\
 & SiO $J=1$--0 $v=1$ &  &  & $<$0.13 & 0.068  & 060215.5 \\
 & SiO $J=1$--0 $v=2$ & 53.0  & 0.42  & 0.81  & 0.072  & 060215.5 \\
 & SiO $J=1$--0 $v=3$ &  &  & $<$0.15 & 0.074  & 060215.5 \\
 & SiO $J=1$--0 $v=4$ &  &  & $<$0.14 & 0.072  & 060215.5 \\
 & $^{29}$SiO $J=1$--0 $v=0$ &  &  & $<$0.12 & 0.061  & 060215.5 \\
 & $^{30}$SiO $J=1$--0 $v=0$ &  &  & $<$0.15 & 0.074  & 060215.5 \\
19067+0811 & SiO $J=1$--0 $v=0$ &  &  & $<$0.14 & 0.071  & 060215.5 \\
(V1368 Aql) & SiO $J=1$--0 $v=1$ & 63.1  & 1.38  & 3.94  & 0.060  & 040511.3 \\
 &  & 61.5  & 0.74  & 1.57  & 0.080  & 060215.5 \\
 & SiO $J=1$--0 $v=2$ & 62.2  & 2.44  & 8.92  & 0.076  & 040511.3 \\
 &  & 62.0  & 0.84  & 2.88  & 0.087  & 060215.5 \\
 & SiO $J=1$--0 $v=3$ & 62.1  & 0.50  & 0.93  & 0.085  & 040515.3 \\
 &  &  &  & $<$0.16 & 0.081  & 060215.5 \\
 & SiO $J=1$--0 $v=4$ &  &  & $<$0.15 & 0.073  & 060215.5 \\
 & SiO $J=2$--1 $v=1$ & 63.3  & 0.33  & 1.11  & 0.048  & 040511.3 \\
 & SiO $J=2$--1 $v=2$ &  &  & $<$0.15 & 0.043  & 040512.3 \\
 & $^{29}$SiO $J=1$--0 $v=0$ & 63.1  & 0.24  & 0.64  & 0.016  & 040512.3 \\
 &  &  &  & $<$0.13 & 0.067  & 060215.5 \\
 & $^{29}$SiO $J=2$--0 $v=0$ & 63.0  & 0.30  & 0.62  & 0.043  & 040512.3 \\
 & $^{30}$SiO $J=1$--0 $v=0$ &  &  & $<$0.14 & 0.070  & 060215.5 \\
 & H$_2$O 6$_{1,6}$$-$5$_{2,3}$ &  &  & $<$0.19 & 0.070  & 040510.3 \\
19081+0322 & SiO $J=1$--0 $v=0$ &  &  & $<$0.14 & 0.072  & 060215.5 \\
 & SiO $J=1$--0 $v=1$ &  &  & $<$0.15 & 0.079  & 060215.5 \\
 & SiO $J=1$--0 $v=2$ & 41.2  & 0.43  & 0.63  & 0.074  & 060215.5 \\
 & SiO $J=1$--0 $v=3$ &  &  & $<$0.16 & 0.080  & 060215.5 \\
 & SiO $J=1$--0 $v=4$ &  &  & $<$0.15 & 0.076  & 060215.5 \\
 & $^{29}$SiO $J=1$--0 $v=0$ &  &  & $<$0.13 & 0.067  & 060215.5 \\
 & $^{30}$SiO $J=1$--0 $v=0$ &  &  & $<$0.15 & 0.078  & 060215.5 \\
19126$-$0708  & SiO $J=1$--0 $v=1$ & $-$21.9  & 8.40  & 10.09  & 0.046  & 040519.2 \\
(W Aql) & SiO $J=1$--0 $v=2$ &  &  & $<$0.12 & 0.059  & 040519.2 \\
 & SiO $J=2$--1 $v=2$ &  &  & $<$0.13 & 0.038  & 040519.2 \\
 & $^{29}$SiO $J=1$--0 $v=0$ &  &  & $<$0.12 & 0.025  & 040519.2 \\
 & $^{29}$SiO $J=2$--0 $v=0$ &  &  & $<$0.14 & 0.039  & 040519.2 \\
19128+0910 & SiO $J=1$--0 $v=0$ &  &  & $<$0.20 & 0.101  & 060217.7 \\
 & SiO $J=1$--0 $v=1$ &  &  & $<$0.20 & 0.103  & 060217.7 \\
 & SiO $J=1$--0 $v=2$ &  &  & $<$0.21 & 0.105  & 060217.7 \\
 & SiO $J=1$--0 $v=3$ &  &  & $<$0.22 & 0.113  & 060217.7 \\
 & SiO $J=1$--0 $v=4$ &  &  & $<$0.21 & 0.108  & 060217.7 \\
 & $^{29}$SiO $J=1$--0 $v=0$ &  &  & $<$0.18 & 0.093  & 060217.7 \\
 & $^{30}$SiO $J=1$--0 $v=0$ &  &  & $<$0.21 & 0.107  & 060217.7 \\
19161+2343 & H$_2$O 6$_{1,6}$$-$5$_{2,3}$ & 15.3  & 12.20  & 20.13  & 0.080  & 040516.2 \\
19192+0922 & SiO $J=1$--0 $v=1$ & $-$68.5  & 9.08  & 19.37  & 0.085  & 040511.3 \\
 & SiO $J=1$--0 $v=2$ & $-$69.4  & 8.92  & 21.67  & 0.122  & 040511.3 \\
 & SiO $J=1$--0 $v=3$ & $-$69.4  & 1.24  & 2.19  & 0.106  & 040515.3 \\
 & SiO $J=2$--1 $v=1$ & $-$69.0  & 1.38  & 4.44  & 0.075  & 040511.3 \\
 & SiO $J=2$--1 $v=2$ &  &  & $<$0.11 & 0.032  & 040512.3 \\
 & $^{29}$SiO $J=1$--0 $v=0$ & $-$70.1  & 0.64  & 1.66  & 0.046  & 040511.3 \\
 & $^{29}$SiO $J=2$--0 $v=0$ &  &  & $<$0.11 & 0.032  & 040512.3 \\
 & H$_2$O 6$_{1,6}$$-$5$_{2,3}$ & $-$58.9  & 2.46  & 8.56  & 0.096  & 040510.2 \\
W51 & SiO $J=1$--0 $v=1$ & 49.2  & 0.67  & 1.55  & 0.065  & 040512.3 \\
 & SiO $J=1$--0 $v=2$ & 49.2  & 0.91  & 2.87  & 0.079  & 040512.2 \\
 & SiO $J=1$--0 $v=3$ &  &  & $<$0.15 & 0.076  & 040515.3 \\
 & SiO $J=2$--1 $v=2$ &  &  & $<$0.21 & 0.060  & 040512.3 \\
 & $^{29}$SiO $J=1$--0 $v=0$ &  &  & $<$0.23 & 0.048  & 040512.3 \\
 & $^{29}$SiO $J=2$--0 $v=0$ &  &  & $<$0.21 & 0.060  & 040512.3 \\
 & H$_2$O 6$_{1,6}$$-$5$_{2,3}$ & 51.9  & 252.44  & 3975.16  & 0.168  & 040516.2 \\
19231+3555 & SiO $J=1$--0 $v=1$ & $-$22.2  & 0.24  & 0.59  & 0.054  & 040514.3 \\
 & SiO $J=1$--0 $v=2$ & $-$25.0  & 1.19  & 2.04  & 0.064  & 040514.3 \\
 & SiO $J=1$--0 $v=3$ &  &  & $<$0.11 & 0.054  & 040515.3 \\
 & SiO $J=2$--1 $v=2$ &  &  & $<$0.15 & 0.044  & 040514.3 \\
 & $^{29}$SiO $J=1$--0 $v=0$ &  &  & $<$0.13 & 0.026  & 040514.3 \\
 & $^{29}$SiO $J=2$--0 $v=0$ &  &  & $<$0.15 & 0.044  & 040514.3 \\
 & H$_2$O 6$_{1,6}$$-$5$_{2,3}$ &  &  & $<$0.13 & 0.048  & 040516.3 \\
19252+2201 & H$_2$O 6$_{1,6}$$-$5$_{2,3}$ &  &  & $<$0.12 & 0.045  & 040516.2 \\
19254+1631 & SiO $J=1$--0 $v=0$ &  &  & $<$0.11 & 0.058  & 060215.5 \\
 & SiO $J=1$--0 $v=1$ & 0.3  & 0.91  & 3.38  & 0.038  & 040511.3 \\
 &  & $-$0.7  & 0.69  & 2.84  & 0.071  & 060215.5 \\
 & SiO $J=1$--0 $v=2$ & 0.0  & 0.89  & 4.72  & 0.045  & 040511.3 \\
 &  & $-$0.1  & 1.00  & 5.14  & 0.096  & 060215.5 \\
 & SiO $J=1$--0 $v=3$ & 3.0  & 0.34  & 0.38  & 0.054  & 040518.3 \\
 &  & 3.1  & 0.31  & 0.34  & 0.058  & 060215.5 \\
 & SiO $J=1$--0 $v=4$ &  &  & $<$0.10 & 0.052  & 060215.5 \\
 & SiO $J=2$--1 $v=1$ & 1.9  & 0.15  & 0.69  & 0.030  & 040511.3 \\
 & $^{29}$SiO $J=1$--0 $v=0$ & 0.7  & 0.12  & 0.62  & 0.019  & 040511.3 \\
 &  & $-$0.2  & 0.16  & 0.53  & 0.047  & 060215.5 \\
 & $^{30}$SiO $J=1$--0 $v=0$ &  &  & $<$0.10 & 0.052  & 060215.5 \\
 & H$_2$O 6$_{1,6}$$-$5$_{2,3}$ &  &  & $<$0.19 & 0.069  & 040510.3 \\
19283+1944 & SiO $J=1$--0 $v=0$ &  &  & $<$0.28 & 0.143  & 060215.5 \\
 & SiO $J=1$--0 $v=1$ &  &  & $<$0.12 & 0.059  & 040514.3 \\
 &  &  &  & $<$0.30 & 0.153  & 060215.5 \\
 & SiO $J=1$--0 $v=2$ & 28.9  & 0.49  & 1.80  & 0.053  & 040520.2 \\
 &  &  &  & $<$0.30 & 0.153  & 060215.5 \\
 & SiO $J=1$--0 $v=3$ & 28.0  & 0.32  & 0.43  & 0.053  & 040520.2 \\
 &  &  &  & $<$0.33 & 0.168  & 060215.5 \\
 & SiO $J=1$--0 $v=4$ &  &  & $<$0.30 & 0.149  & 060215.5 \\
 & SiO $J=2$--1 $v=2$ &  &  & $<$0.17 & 0.048  & 040514.3 \\
 & $^{29}$SiO $J=1$--0 $v=0$ &  &  & $<$0.14 & 0.029  & 040514.3 \\
 &  &  &  & $<$0.26 & 0.131  & 060215.5 \\
 & $^{29}$SiO $J=2$--0 $v=0$ &  &  & $<$0.17 & 0.048  & 040514.3 \\
 & $^{30}$SiO $J=1$--0 $v=0$ &  &  & $<$0.29 & 0.150  & 060215.5 \\
19295+2228 & SiO $J=1$--0 $v=0$ &  &  & $<$0.09 & 0.048  & 060215.6 \\
 & SiO $J=1$--0 $v=1$ & $-$72.4  & 0.77  & 1.81  & 0.063  & 060215.6 \\
 & SiO $J=1$--0 $v=2$ & $-$72.4  & 1.04  & 3.33  & 0.080  & 060215.6 \\
 & SiO $J=1$--0 $v=3$ &  &  & $<$0.11 & 0.054  & 060215.6 \\
 & SiO $J=1$--0 $v=4$ &  &  & $<$0.10 & 0.050  & 060215.6 \\
 & $^{29}$SiO $J=1$--0 $v=0$ &  &  & $<$0.09 & 0.046  & 060215.6 \\
 & $^{30}$SiO $J=1$--0 $v=0$ &  &  & $<$0.10 & 0.052  & 060215.6 \\
19312+1950 & SiO $J=1$--0 $v=1$ & 50.6  & 0.18  & 0.67  & 0.039  & 040514.3 \\
 & SiO $J=1$--0 $v=2$ & 35.6  & 0.40  & 1.77  & 0.048  & 040514.3 \\
 & SiO $J=1$--0 $v=3$ &  &  & $<$0.08 & 0.042  & 040515.3 \\
 & SiO $J=2$--1 $v=2$ &  &  & $<$0.12 & 0.034  & 040514.3 \\
 & $^{29}$SiO $J=1$--0 $v=0$ &  &  & $<$0.09 & 0.019  & 040514.3 \\
 & $^{29}$SiO $J=2$--0 $v=0$ &  &  & $<$0.12 & 0.034  & 040514.3 \\
 & H$_2$O 6$_{1,6}$$-$5$_{2,3}$ & 33.3  & 4.36  & 13.53  & 0.059  & 040510.3 \\
19354+5005 & SiO $J=1$--0 $v=1$ &  &  & $<$0.28 & 0.141  & 040519.3 \\
(R Cyg) & SiO $J=1$--0 $v=2$ &  &  & $<$0.36 & 0.184  & 040519.3 \\
 & SiO $J=2$--1 $v=2$ &  &  & $<$0.41 & 0.118  & 040519.3 \\
 & $^{29}$SiO $J=1$--0 $v=0$ &  &  & $<$0.38 & 0.077  & 040519.3 \\
 & $^{29}$SiO $J=2$--0 $v=0$ &  &  & $<$0.40 & 0.117  & 040519.3 \\
19374+0550 & SiO $J=1$--0 $v=2$ & $-$16.8  & 0.77  & 2.75  & 0.096  & 040520.3 \\
 & SiO $J=1$--0 $v=3$ &  &  & $<$0.19 & 0.098  & 040520.3 \\
 & $^{29}$SiO $J=1$--0 $v=0$ & $-$16.6  & 0.15  & 0.40  & 0.035  & 040520.3 \\
 & H$_2$O 6$_{1,6}$$-$5$_{2,3}$ &  &  & $<$0.23 & 0.085  & 040510.3 \\
19422+3506 & SiO $J=1$--0 $v=1$ & $-$47.0  & 3.10  & 10.27  & 0.062  & 040511.3 \\
 & SiO $J=1$--0 $v=2$ & $-$46.5  & 4.79  & 14.31  & 0.081  & 040511.3 \\
 & SiO $J=1$--0 $v=3$ & $-$48.0  & 0.55  & 1.30  & 0.074  & 040515.3 \\
 & SiO $J=2$--1 $v=1$ & $-$51.0  & 0.83  & 4.32  & 0.052  & 040511.3 \\
 & SiO $J=2$--1 $v=2$ &  &  & $<$0.10 & 0.029  & 040514.3 \\
 & $^{29}$SiO $J=1$--0 $v=0$ &  &  & $<$0.09 & 0.018  & 040514.3 \\
 & $^{29}$SiO $J=2$--0 $v=0$ &  &  & $<$0.10 & 0.029  & 040514.3 \\
 & H$_2$O 6$_{1,6}$$-$5$_{2,3}$ & $-$48.9  & 1.43  & 12.41  & 0.076  & 040510.3 \\
19440+2251 & SiO $J=1$--0 $v=0$ &  &  & $<$0.17 & 0.090  & 060217.5 \\
 & SiO $J=1$--0 $v=1$ & $-$8.2  & 0.59  & 1.21  & 0.100  & 060217.5 \\
 & SiO $J=1$--0 $v=2$ & $-$8.2  & 0.82  & 1.75  & 0.105  & 060217.5 \\
 & SiO $J=1$--0 $v=3$ &  &  & $<$0.21 & 0.108  & 060217.5 \\
 & SiO $J=1$--0 $v=4$ &  &  & $<$0.20 & 0.103  & 060217.5 \\
 & $^{29}$SiO $J=1$--0 $v=0$ &  &  & $<$0.16 & 0.083  & 060217.5 \\
 & $^{30}$SiO $J=1$--0 $v=0$ &  &  & $<$0.19 & 0.096  & 060217.5 \\
19486+3247  & SiO $J=1$--0 $v=0$ & 9.4  & 0.83  & 4.76  & 0.075  & 060217.5 \\
($\chi$ Cyg) & SiO $J=1$--0 $v=1$ & 9.4  & 3.67  & 6.90  & 0.066  & 040514.3 \\
 &  & 8.6  & 7.88  & 27.12  & 0.079  & 060217.5 \\
 & SiO $J=1$--0 $v=2$ &  &  & $<$0.15 & 0.078  & 040514.3 \\
 &  & 7.9  & 11.32  & 9.88  & 0.075  & 060217.5 \\
 & SiO $J=1$--0 $v=3$ &  &  & $<$0.19 & 0.094  & 040518.3 \\
 &  &  &  & $<$0.17 & 0.085  & 060217.5 \\
 & SiO $J=1$--0 $v=4$ &  &  & $<$0.16 & 0.080  & 060217.5 \\
 & SiO $J=2$--1 $v=2$ & 10.7  & 1.15  & 3.33  & 0.059  & 040514.3 \\
 & $^{29}$SiO $J=1$--0 $v=0$ & 8.7  & 0.38  & 0.67  & 0.070  & 060217.5 \\
 & $^{29}$SiO $J=2$--0 $v=0$ & 7.2  & 0.24  & 1.64  & 0.059  & 040514.3 \\
 & $^{30}$SiO $J=1$--0 $v=0$ &  &  & $<$0.16 & 0.082  & 060217.5 \\
19493+2905 & SiO $J=1$--0 $v=1$ & $-$20.5  & 1.26  & 4.90  & 0.047  & 040511.3 \\
 & SiO $J=1$--0 $v=2$ & $-$21.2  & 1.30  & 5.50  & 0.061  & 040511.3 \\
 & SiO $J=1$--0 $v=3$ &  &  & $<$0.08 & 0.040  & 040515.3 \\
 & SiO $J=2$--1 $v=1$ & $-$22.2  & 0.28  & 1.37  & 0.040  & 040511.3 \\
 & SiO $J=2$--1 $v=2$ &  &  & $<$0.19 & 0.054  & 040519.3 \\
 & $^{29}$SiO $J=1$--0 $v=0$ &  &  & $<$0.12 & 0.024  & 040511.3 \\
 & $^{29}$SiO $J=2$--0 $v=0$ &  &  & $<$0.19 & 0.055  & 040519.3 \\
 & H$_2$O 6$_{1,6}$$-$5$_{2,3}$ &  &  & $<$0.19 & 0.070  & 040510.3 \\
19508+2659 & H$_2$O 6$_{1,6}$$-$5$_{2,3}$ & 6.3  & 0.42  & 2.75  & 0.083  & 040516.3 \\
19576+2814 & SiO $J=1$--0 $v=0$ &  &  & $<$0.18 & 0.090  & 060217.5 \\
 & SiO $J=1$--0 $v=1$ &  &  & $<$0.19 & 0.095  & 060217.5 \\
 & SiO $J=1$--0 $v=2$ &  &  & $<$0.19 & 0.094  & 060217.5 \\
 & SiO $J=1$--0 $v=3$ &  &  & $<$0.22 & 0.109  & 060217.5 \\
 & SiO $J=1$--0 $v=4$ &  &  & $<$0.20 & 0.099  & 060217.5 \\
 & $^{29}$SiO $J=1$--0 $v=0$ &  &  & $<$0.17 & 0.084  & 060217.5 \\
 & $^{30}$SiO $J=1$--0 $v=0$ &  &  & $<$0.20 & 0.099  & 060217.5 \\
 & H$_2$O 6$_{1,6}$$-$5$_{2,3}$ &  &  & $<$0.35 & 0.131  & 040516.3 \\
20010+3011 & H$_2$O 6$_{1,6}$$-$5$_{2,3}$ & 23.3  & 3.32  & 11.24  & 0.071  & 040516.3 \\
20024+1727 & SiO $J=1$--0 $v=1$ & $-$1.9  & 8.74  & 18.61  & 0.065  & 040519.3 \\
(V718 Cyg) & SiO $J=1$--0 $v=2$ & $-$0.9  & 6.17  & 12.06  & 0.083  & 040519.3 \\
 & SiO $J=1$--0 $v=3$ &  &  & $<$0.10 & 0.050  & 040518.3 \\
 & SiO $J=2$--1 $v=2$ &  &  & $<$0.18 & 0.052  & 040519.3 \\
 & $^{29}$SiO $J=1$--0 $v=0$ & $-$3.0  & 0.13  & 0.42  & 0.034  & 040519.3 \\
 & $^{29}$SiO $J=2$--0 $v=0$ &  &  & $<$0.17 & 0.049  & 040519.3 \\
 & H$_2$O 6$_{1,6}$$-$5$_{2,3}$ &  &  & $<$0.24 & 0.090  & 040510.3 \\
20043+2653 & SiO $J=1$--0 $v=0$ &  &  & $<$0.13 & 0.065  & 060217.5 \\
 & SiO $J=1$--0 $v=1$ & $-$4.3  & 0.76  & 1.41  & 0.066  & 040519.3 \\
 &  & $-$4.8  & 0.43  & 1.09  & 0.075  & 060217.5 \\
 & SiO $J=1$--0 $v=2$ & $-$4.9  & 1.24  & 3.60  & 0.090  & 040520.3 \\
 &  & $-$5.1  & 0.93  & 1.97  & 0.081  & 060217.5 \\
 & SiO $J=1$--0 $v=3$ & $-$4.8  & 0.72  & 1.35  & 0.096  & 040520.3 \\
 &  & $-$7.3  & 0.24  & 0.36  & 0.076  & 060217.5 \\
 & SiO $J=1$--0 $v=4$ &  &  & $<$0.14 & 0.070  & 060217.5 \\
 & SiO $J=2$--1 $v=2$ &  &  & $<$0.18 & 0.053  & 040519.3 \\
 & $^{29}$SiO $J=1$--0 $v=0$ &  &  & $<$0.17 & 0.035  & 040519.3 \\
 &  &  &  & $<$0.12 & 0.060  & 060217.5 \\
 & $^{29}$SiO $J=2$--0 $v=0$ &  &  & $<$0.19 & 0.054  & 040519.3 \\
 & $^{30}$SiO $J=1$--0 $v=0$ &  &  & $<$0.14 & 0.070  & 060217.5 \\
 & H$_2$O 6$_{1,6}$$-$5$_{2,3}$ &  &  & $<$0.20 & 0.074  & 040516.3 \\
20052+0554 & SiO $J=1$--0 $v=0$ &  &  & $<$0.15 & 0.077  & 060217.6 \\
(V1416 Aql) & SiO $J=1$--0 $v=1$ & $-$67.2  & 4.85  & 21.47  & 0.082  & 060217.6 \\
 & SiO $J=1$--0 $v=2$ & $-$66.6  & 4.02  & 19.74  & 0.083  & 060217.6 \\
 & SiO $J=1$--0 $v=3$ & $-$71.8  & 0.60  & 0.53  & 0.094  & 060217.6 \\
 & SiO $J=1$--0 $v=4$ &  &  & $<$0.18 & 0.089  & 060217.6 \\
 & $^{29}$SiO $J=1$--0 $v=0$ & $-$70.4  & 0.33  & 0.50  & 0.077  & 060217.6 \\
 & $^{30}$SiO $J=1$--0 $v=0$ &  &  & $<$0.17 & 0.088  & 060217.6 \\
20077$-$0625 & SiO $J=1$--0 $v=0$ & $-$14.3 & 0.40  & 3.16  & 0.072  & 060217.6 \\
(V1300 Aql) & SiO $J=1$--0 $v=1$ & $-$17.5  & 2.62  & 8.48  & 0.080  & 060217.6 \\
 & SiO $J=1$--0 $v=2$ & $-$17.1  & 5.91  & 14.99  & 0.074  & 060217.6 \\
 & SiO $J=1$--0 $v=3$ &  &  & $<$0.16 & 0.082  & 060217.6 \\
 & SiO $J=1$--0 $v=4$ &  &  & $<$0.15 & 0.077  & 060217.6 \\
 & $^{29}$SiO $J=1$--0 $v=0$ & $-$18.6  & 0.39  & 0.57  & 0.069  & 060217.6 \\
 & $^{30}$SiO $J=1$--0 $v=0$ &  &  & $<$0.15 & 0.078  & 060217.6 \\
20095+2726 & H$_2$O 6$_{1,6}$$-$5$_{2,3}$ &  &  & $<$0.16 & 0.058  & 040516.3 \\
20241+3811 & SiO $J=1$--0 $v=0$ &  &  & $<$0.14 & 0.072  & 060218.5 \\
(KY Cyg) & SiO $J=1$--0 $v=1$ & $-$0.5  & 1.29  & 5.13  & 0.117  & 060218.5 \\
 & SiO $J=1$--0 $v=2$ & $-$0.6  & 0.84  & 1.48  & 0.087  & 060218.5 \\
 & SiO $J=1$--0 $v=3$ &  &  & $<$0.18 & 0.089  & 060218.5 \\
 & SiO $J=1$--0 $v=4$ &  &  & $<$0.16 & 0.080  & 060218.5 \\
 & $^{29}$SiO $J=1$--0 $v=0$ &  &  & $<$0.14 & 0.072  & 060218.5 \\
 & $^{30}$SiO $J=1$--0 $v=0$ &  &  & $<$0.16 & 0.081  & 060218.5 \\
20381+5001 & SiO $J=1$--0 $v=2$ & $-$37.5  & 3.28  & 15.16  & 0.142  & 040520.3 \\
 & SiO $J=1$--0 $v=3$ &  &  & $<$0.27 & 0.137  & 040520.3 \\
 & $^{29}$SiO $J=1$--0 $v=0$ &  &  & $<$2.02 & 0.413  & 040520.3 \\
20459+5015 & SiO $J=1$--0 $v=2$ & $-$34.5  & 0.82  & 1.29  & 0.094  & 040520.3 \\
 & SiO $J=1$--0 $v=3$ &  &  & $<$0.19 & 0.095  & 040520.3 \\
20491+4236 & SiO $J=1$--0 $v=0$ &  &  & $<$0.15 & 0.076  & 060217.5 \\
 & SiO $J=1$--0 $v=1$ & $-$40.6  & 1.80  & 4.87  & 0.046  & 040511.3 \\
 &  & $-$40.2  & 3.45  & 9.91  & 0.083  & 060217.5 \\
 & SiO $J=1$--0 $v=2$ & $-$41.9  & 2.54  & 7.83  & 0.060  & 040511.3 \\
 &  & $-$41.2  & 4.51  & 13.67  & 0.079  & 060217.5 \\
 & SiO $J=1$--0 $v=3$ & $-$40.8  & 0.31  & 0.42  & 0.068  & 040520.3 \\
 &  & $-$40.4  & 0.85  & 1.54  & 0.090  & 060217.5 \\
 & SiO $J=1$--0 $v=4$ &  &  & $<$0.16 & 0.082  & 060217.5 \\
 & SiO $J=2$--1 $v=1$ & $-$40.7  & 0.47  & 1.59  & 0.037  & 040511.3 \\
 & $^{29}$SiO $J=1$--0 $v=0$ &  &  & $<$0.11 & 0.023  & 040511.3 \\
 &  &  &  & $<$0.14 & 0.073  & 060217.5 \\
 & $^{30}$SiO $J=1$--0 $v=0$ &  &  & $<$0.16 & 0.079  & 060217.5 \\
 & H$_2$O 6$_{1,6}$$-$5$_{2,3}$ & $-$38.7  & 0.80  & 2.64  & 0.056  & 040510.3 \\
20529+3013 & SiO $J=1$--0 $v=0$ &  &  & $<$0.12 & 0.060  & 060218.5 \\
(UX Cyg) & SiO $J=1$--0 $v=1$ & 1.6  & 3.52  & 19.74  & 0.063  & 060218.5 \\
 & SiO $J=1$--0 $v=2$ & 1.5  & 3.65  & 19.62  & 0.059  & 060218.5 \\
 & SiO $J=1$--0 $v=3$ & 0.1  & 0.64  & 0.70  & 0.070  & 060218.5 \\
 & SiO $J=1$--0 $v=4$ &  &  & $<$0.12 & 0.061  & 060218.5 \\
 & $^{29}$SiO $J=1$--0 $v=0$ & 2.9  & 0.34  & 0.67  & 0.054  & 060218.5 \\
 & $^{30}$SiO $J=1$--0 $v=0$ &  &  & $<$0.13 & 0.067  & 060218.5 \\
21088+6817 & SiO $J=1$--0 $v=0$ &  &  & $<$0.11 & 0.055  & 060219.5 \\
(T Cep) & SiO $J=1$--0 $v=1$ & $-$3.2  & 36.77  & 116.63  & 0.058  & 060219.5 \\
 & SiO $J=1$--0 $v=2$ & $-$2.6  & 42.27  & 126.89  & 0.058  & 060219.5 \\
 & SiO $J=1$--0 $v=3$ &  &  & $<$0.13 & 0.064  & 060219.5 \\
 & SiO $J=1$--0 $v=4$ &  &  & $<$0.11 & 0.058  & 060219.5 \\
 & $^{29}$SiO $J=1$--0 $v=0$ &  &  & $<$0.10 & 0.052  & 060219.5 \\
 & $^{30}$SiO $J=1$--0 $v=0$ &  &  & $<$0.12 & 0.058  & 060219.5 \\
21270+7135 & SiO $J=1$--0 $v=0$ &  &  & $<$0.11 & 0.055  & 060219.5 \\
(V363 Cep) & SiO $J=1$--0 $v=1$ & $-$34.5  & 3.59  & 14.53  & 0.059  & 060219.5 \\
 & SiO $J=1$--0 $v=2$ & $-$34.6  & 2.61  & 12.20  & 0.054  & 060219.5 \\
 & SiO $J=1$--0 $v=3$ & $-$31.9  & 0.50  & 1.26  & 0.065  & 060219.5 \\
 & SiO $J=1$--0 $v=4$ &  &  & $<$0.12 & 0.059  & 060219.5 \\
 & $^{29}$SiO $J=1$--0 $v=0$ & $-$31.8  & 0.39  & 1.00  & 0.052  & 060219.5 \\
 & $^{30}$SiO $J=1$--0 $v=0$ &  &  & $<$0.12 & 0.059  & 060219.5 \\
21286+1055 & SiO $J=1$--0 $v=0$ &  &  & $<$0.15 & 0.079  & 060218.5 \\
(UU Peg) & SiO $J=1$--0 $v=1$ & 33.5  & 5.63  & 23.01  & 0.089  & 060218.5 \\
 & SiO $J=1$--0 $v=2$ & 29.9  & 7.19  & 21.35  & 0.085  & 060218.5 \\
 & SiO $J=1$--0 $v=3$ &  &  & $<$0.19 & 0.096  & 060218.5 \\
 & SiO $J=1$--0 $v=4$ &  &  & $<$0.17 & 0.086  & 060218.5 \\
 & $^{29}$SiO $J=1$--0 $v=0$ & 29.8  & 0.27  & 0.53  & 0.077  & 060218.5 \\
 & $^{30}$SiO $J=1$--0 $v=0$ &  &  & $<$0.17 & 0.088  & 060218.5 \\
21419+5832 & SiO $J=1$--0 $v=0$ &  &  & $<$0.18 & 0.090  & 060219.5 \\
($\mu$ Cep) & SiO $J=1$--0 $v=1$ & 27.9  & 1.62  & 3.17  & 0.096  & 060219.5 \\
 & SiO $J=1$--0 $v=2$ & 27.5  & 6.01  & 13.86  & 0.092  & 060219.5 \\
 & SiO $J=1$--0 $v=3$ & 27.5  & 1.61  & 3.77  & 0.106  & 060219.5 \\
 & SiO $J=1$--0 $v=4$ &  &  & $<$0.18 & 0.093  & 060219.5 \\
 & $^{29}$SiO $J=1$--0 $v=0$ &  &  & $<$0.17 & 0.084  & 060219.5 \\
 & $^{30}$SiO $J=1$--0 $v=0$ &  &  & $<$0.19 & 0.095  & 060219.5 \\
21426+1228 & SiO $J=1$--0 $v=0$ &  &  & $<$0.17 & 0.088  & 060218.6 \\
(TU Peg) & SiO $J=1$--0 $v=1$ & 12.5  & 12.45  & 31.78  & 0.097  & 060218.6 \\
 & SiO $J=1$--0 $v=2$ & 12.3  & 8.07  & 22.84  & 0.093  & 060218.6 \\
 & SiO $J=1$--0 $v=3$ &  &  & $<$0.21 & 0.107  & 060218.6 \\
 & SiO $J=1$--0 $v=4$ &  &  & $<$0.19 & 0.098  & 060218.6 \\
 & $^{29}$SiO $J=1$--0 $v=0$ &  &  & $<$0.16 & 0.082  & 060218.6 \\
 & $^{30}$SiO $J=1$--0 $v=0$ &  &  & $<$0.19 & 0.097  & 060218.6 \\
21439$-$0226  & H$_2$O 6$_{1,6}$$-$5$_{2,3}$ &  &  & $<$0.23 & 0.085  & 040517.3 \\
(EP Aqr) &  &  &  &  &  &  \\
21456+6422 & SiO $J=1$--0 $v=0$ &  &  & $<$0.18 & 0.092  & 060219.6 \\
(RT Cep) & SiO $J=1$--0 $v=1$ & $-$43.6  & 5.60  & 13.74  & 0.104  & 060219.6 \\
 & SiO $J=1$--0 $v=2$ & $-$44.5  & 9.36  & 21.45  & 0.092  & 060219.6 \\
 & SiO $J=1$--0 $v=3$ &  &  & $<$0.21 & 0.108  & 060219.6 \\
 & SiO $J=1$--0 $v=4$ &  &  & $<$0.19 & 0.093  & 060219.6 \\
 & $^{29}$SiO $J=1$--0 $v=0$ & $-$44.2  & 1.99  & 2.78  & 0.086  & 060219.6 \\
 & $^{30}$SiO $J=1$--0 $v=0$ &  &  & $<$0.20 & 0.099  & 060219.6 \\
22097+5647 & SiO $J=1$--0 $v=0$ &  &  & $<$0.17 & 0.088  & 060219.6 \\
(CU Cep) & SiO $J=1$--0 $v=1$ & $-$49.2  & 4.84  & 28.29  & 0.098  & 060219.6 \\
 & SiO $J=1$--0 $v=2$ & $-$48.5  & 2.90  & 10.13  & 0.090  & 060219.6 \\
 & SiO $J=1$--0 $v=3$ &  &  & $<$0.21 & 0.107  & 060219.6 \\
 & SiO $J=1$--0 $v=4$ &  &  & $<$0.19 & 0.095  & 060219.6 \\
 & $^{29}$SiO $J=1$--0 $v=0$ &  &  & $<$0.18 & 0.091  & 060219.6 \\
 & $^{30}$SiO $J=1$--0 $v=0$ &  &  & $<$0.19 & 0.096  & 060219.6 \\
22177+5936 & SiO $J=1$--0 $v=0$ &  &  & $<$0.22 & 0.115  & 060219.6 \\
 & SiO $J=1$--0 $v=1$ & $-$27.2  & 1.05  & 2.19  & 0.127  & 060219.6 \\
 & SiO $J=1$--0 $v=2$ & $-$27.4  & 9.53  & 22.58  & 0.127  & 060219.6 \\
 & SiO $J=1$--0 $v=3$ &  &  & $<$0.27 & 0.138  & 060219.6 \\
 & SiO $J=1$--0 $v=4$ &  &  & $<$0.24 & 0.123  & 060219.6 \\
 & $^{29}$SiO $J=1$--0 $v=0$ & $-$27.9  & 0.68  & 0.68  & 0.118  & 060219.6 \\
 & $^{30}$SiO $J=1$--0 $v=0$ &  &  & $<$0.28 & 0.140  & 060219.6 \\
22480+6002 & SiO $J=1$--0 $v=2$ & $-$51.2  & 4.41  & 19.43  & 0.110  & 040520.3 \\
 & SiO $J=1$--0 $v=3$ &  &  & $<$0.21 & 0.109  & 040520.3 \\
 & $^{29}$SiO $J=1$--0 $v=0$ &  &  & $<$1.25 & 0.257  & 040520.3 \\
 & H$_2$O 6$_{1,6}$$-$5$_{2,3}$ & $-$50.8  & 4.94  & 30.35  & 0.168  & 040516.3 \\
22512+6100 & SiO $J=1$--0 $v=0$ &  &  & $<$0.20 & 0.102  & 060220.4 \\
(V386 Cep) & SiO $J=1$--0 $v=1$ & $-$50.4  & 1.29  & 9.66  & 0.104  & 060220.4 \\
 & SiO $J=1$--0 $v=2$ & $-$51.6  & 1.25  & 6.70  & 0.103  & 060220.4 \\
 & SiO $J=1$--0 $v=3$ &  &  & $<$0.24 & 0.119  & 060220.4 \\
 & SiO $J=1$--0 $v=4$ &  &  & $<$0.21 & 0.105  & 060220.4 \\
 & $^{29}$SiO $J=1$--0 $v=0$ &  &  & $<$0.19 & 0.095  & 060220.4 \\
 & $^{30}$SiO $J=1$--0 $v=0$ &  &  & $<$0.22 & 0.111  & 060220.4 \\
22516+0838 & SiO $J=1$--0 $v=0$ &  &  & $<$0.19 & 0.097  & 060218.6 \\
(KZ Peg) & SiO $J=1$--0 $v=1$ & 5.0  & 9.26  & 33.16  & 0.107  & 060218.6 \\
 & SiO $J=1$--0 $v=2$ & 3.0  & 10.93  & 30.25  & 0.103  & 060218.6 \\
 & SiO $J=1$--0 $v=3$ & 1.6  & 1.40  & 1.70  & 0.119  & 060218.6 \\
 & SiO $J=1$--0 $v=4$ &  &  & $<$0.22 & 0.111  & 060218.6 \\
 & $^{29}$SiO $J=1$--0 $v=0$ &  &  & $<$0.18 & 0.092  & 060218.6 \\
 & $^{30}$SiO $J=1$--0 $v=0$ &  &  & $<$0.22 & 0.113  & 060218.6 \\
22525+6033 & SiO $J=1$--0 $v=0$ &  &  & $<$0.20 & 0.101  & 060220.5 \\
(MY Cep) & SiO $J=1$--0 $v=1$ & $-$53.7  & 3.77  & 17.08  & 0.110  & 060220.5 \\
 & SiO $J=1$--0 $v=2$ & $-$51.9  & 2.15  & 5.56  & 0.106  & 060220.5 \\
 & SiO $J=1$--0 $v=3$ &  &  & $<$0.24 & 0.119  & 060220.5 \\
 & SiO $J=1$--0 $v=4$ &  &  & $<$0.24 & 0.121  & 060220.5 \\
 & $^{29}$SiO $J=1$--0 $v=0$ &  &  & $<$0.20 & 0.103  & 060220.5 \\
 & $^{30}$SiO $J=1$--0 $v=0$ &  &  & $<$0.21 & 0.108  & 060220.5 \\
23041+1016 & SiO $J=1$--0 $v=0$ &  &  & $<$0.17 & 0.087  & 060218.6 \\
(R Peg) & SiO $J=1$--0 $v=1$ & 26.6  & 2.85  & 18.66  & 0.095  & 060218.6 \\
 & SiO $J=1$--0 $v=2$ & 27.3  & 2.16  & 14.28  & 0.092  & 060218.6 \\
 & SiO $J=1$--0 $v=3$ &  &  & $<$0.20 & 0.103  & 060218.6 \\
 & SiO $J=1$--0 $v=4$ &  &  & $<$0.19 & 0.097  & 060218.6 \\
 & $^{29}$SiO $J=1$--0 $v=0$ &  &  & $<$0.16 & 0.082  & 060218.6 \\
 & $^{30}$SiO $J=1$--0 $v=0$ &  &  & $<$0.21 & 0.107  & 060218.6 \\
\enddata
\end{deluxetable}

\end{document}